\documentclass[12pt]{article}

\usepackage{graphicx}% Include figure files
\usepackage{amsmath}
\usepackage{amssymb}
\usepackage{amscd}
\usepackage{latexsym}

\usepackage{cite}
\numberwithin{equation}{section}

\newcommand{\be}{\begin{equation}}
\newcommand{\ee}{\end{equation}}
\newcommand{\br}{{\bf r}}
\newcommand{\bt}{\beta}
\newcommand{\vp}{\varphi}
\newcommand{\om}{\omega}
\newcommand{\ep}{\varepsilon}
\newcommand{\lbd}{\lambda}
\newcommand{\al}{\alpha}
\newcommand{\gm}{\gamma}
\newcommand{\Gm}{\Gamma}
\newcommand{\dlt}{\delta}
\newcommand{\cH}{{\cal H}}
\newcommand{\rgl}{\rangle}
\newcommand{\lgl}{\langle}
\newcommand{\ra}{\rightarrow}
\newcommand*{\pr}[1]{\mathcal{#1}}

\begin{document}

\begin{center}

{\Large{\bf
Entanglement production by statistical operators} \\ [5mm]

V.I. Yukalov$^{1,2}$, E.P. Yukalova$^3$, and V.A. Yurovsky$^4$} \\ [3mm]

{\it
$^1$Bogolubov Laboratory of Theoretical Physics, \\
Joint Institute for Nuclear Research, Dubna 141980, Russia \\ [2mm]
                                      
$^2$Instituto de Fisica de S\~{a}o Carlos, Universidade de S\~{a}o Paulo, CP 369,  \\
S\~{a}o Carlos 13560-970, S\~{a}o Paulo, Brazil  \\ [2mm]

$^3$Laboratory of Information Technologies, \\
Joint Institute for Nuclear Research, Dubna 141980, Russia \\ [2mm]

$^4$School of Chemistry, Tel Aviv University, 6997801 Tel Aviv, Israel} \\ [5mm]

{\bf E-mails}: yukalov@theor.jinr.ru, yukalova@theor.jinr.ru, 
volodia@post.tau.ac.il

\end{center}


\vskip 2cm

\begin{abstract}
In the problem of entanglement there exist two different notions. One is the entanglement 
of a quantum state, characterizing the state structure. The other is entanglement 
production by quantum operators, describing the action of operators in the given 
Hilbert space. Entanglement production by statistical operators, or density operators, 
is an important notion arising in quantum measurements and quantum information 
processing. The operational meaning of the entangling power of any operator, including
statistical operators, is the property of the operators to entangle wave functions of the
Hilbert space they are defined on. The measure of entanglement production by statistical
operators is described and illustrated by entangled quantum states, equilibrium Gibbs states, 
as well as by the state of a complex multiparticle spinor system. It is shown that this measure 
is in intimate relation to other notions of quantum information theory, such as the purity of 
quantum states, linear entropy, or impurity, inverse participation ratio, quadratic R\'{e}nyi 
entropy, the correlation function of composite measurements, and decoherence phenomenon.
This measure can be introduced for a set of statistical operators characterizing a system after 
quantum measurements. The explicit value of the measure depends on the type of the Hilbert 
space partitioning. For a general multiparticle spinor system, it is possible to accomplish the 
particle-particle partitioning or spin-spatial partitioning. Conditions are defined showing when 
entanglement production is maximal and when it is zero. The study on entanglement
production by statistical operators is important because, depending on whether such an 
operator is entangling or not, it generates qualitatively different probability measures,
which is principal for quantum measurements and  quantum information processing.    
\end{abstract}

\section{Introduction}

Entanglement is a principally important notion for several branches of quantum
theory, such as quantum measurements, quantum information processing, quantum
computing, and quantum decision theory (see books and reviews
\cite{Williams_1,Nielsen_2,Vedral_3,Keyl_4,Horodecki_76,Guhne_77,Wilde_5,Siewert_78,Yukalov_6}).  
It is possible to distinguish three directions in studying entanglement for composite 
systems described in terms of tensor products of Hilbert spaces.  

One is the entanglement of quantum states, characterized by wave functions in
the case of pure states and by statistical operators, for mixed states. A wave
function is entangled when it cannot be represented by a tensor product of wave
functions pertaining to different Hilbert spaces. And the wave function is disentangled,
when it can be represented as a product
\be
\label{A1}
   \varphi_{dis} = \bigotimes_i \varphi_i \; .
\ee
A statistical operator is entangled if it cannot be represented as a linear combination 
of products of partial statistical operators acting in different Hilbert spaces 
\cite{Williams_1,Nielsen_2,Vedral_3,Keyl_4,Horodecki_76,Guhne_77,Wilde_5,Siewert_78}. 
And it is called separable, if it can be represented as a finite linear combination
\be
\label{A2}
 \hat\rho_{sep} = \sum_k \; p_k \bigotimes_i \hat\rho_{ik} \;  ,
\ee
in which
\be
\label{A3}
0 \leq p_k \leq 1 \; , \qquad \sum_k p_k = 1 \; ,
\ee
and $\hat{\rho}_{ik}$ are statistical operators acting on partial Hilbert spaces
\cite{Williams_1,Nielsen_2,Vedral_3,Keyl_4,Horodecki_76,Guhne_77,Wilde_5}.

The notion of states can be straightforwardly extended to a set of bounded 
operators, which, being complimented by the Hilbert-Schmidt scalar product, forms 
a Hilbert-Schmidt space, where the operators are isomorphic to states of this
space. Then it is admissible to consider the entanglement of operators in a way
similar to the entanglement of states, thus just lifting the notion of entanglement 
from the state level to the operator level 
\cite{Yukalov_6,Zanardi_7,Balakrishnan_8,Macchiavello_9,Yukalov_10,Kong_11}. 

A rather separate problem is the study of entangling properties of unitary operators 
acting on a set of given states. This can be characterized by considering the 
entanglement of states generated by these unitary operators acting on disentangled 
states. In the case of several states, one averages the appropriate measure of a 
unitary operator, say linear entropy, over a set of states with a given distribution 
\cite{Zanardi_7,Macchiavello_9,Kong_11,Zanardi_12}.  
One usually considers unitary operators, since information-processing gates are 
characterized by such operators. In that approach, the problem is reduced to the
consideration of the entanglement of the states, obtained by the action of a unitary 
operator, under the given set of initial states. But this does not describe the 
entangling properties of an operator acting on the whole Hilbert space. 

In the present paper, we consider the related problem of describing entangling 
properties of operators. An operator is called entangling, if there exists at least 
one separable pure state such that it becomes entangled under the action of the 
operator. Conversely, one says that an operator preserves separability if its action 
on any separable pure state yields again a separable pure state. It has been proved 
\cite{Marcus_13,Westwick_14,Johnston_15} that the only operators preserving 
separability are the operators having the form of tensor products of local operators 
and a swap operator permuting Hilbert spaces in the tensor product describing the 
total Hilbert space of a composite system. The action of the swap operator is trivial, 
in the sense that it merely permutes the indices labeling the spaces. This result of 
separability preservation by product operators has been proved for binary 
\cite{Marcus_13,Beasley_16,Alfsen_17} as well as for multipartite systems 
\cite{Westwick_14,Johnston_15,Friedland_18}. The operators preserving separability 
can be called nonentangling \cite{Gohberg_19,Crouzeux_20}. While an operator 
transforming at least one disentangled state into an entangled state is termed 
entangling \cite{Fan_21,Ming_22}. The strongest type of an entangling  operator is 
a universal entangling gate that makes all disentangled pure states entangled \cite{Chen_23}.

The general problem is what could be a measure of entanglement production 
characterizing the entangling properties of an arbitrary operator defined on the whole
Hilbert space of a composite system, but not only for some selected initial states 
from this space. Such a global measure of entanglement production by an arbitrary 
operator has been proposed in Refs. \cite{Yukalov_24,Yukalov_25}. This measure
is applicable to any system, whether bipartite or multipartite, and to any
trace-class operator \cite{Rudin_26,Conway_27}, which does not necessarily need 
to be unitary. The entanglement production has been investigated for several 
physical systems, such as multimode Bose-Einstein condensates of atoms in traps
and in optical lattices \cite{Yukalov_28,Yukalov_29,Yukalov_30} and radiating
resonant atoms \cite{Yukalov_31}. The entanglement production by evolution 
operators has also been studied \cite{Yukalov_32}.  

As is mentioned above, the operator entanglement is usually considered for 
unitary operators, since the evolution operators as well as various information 
gates are unitary. However, it may happen important to study the entanglement 
production by nonunitary operators. For example, one may need to quantify the
entanglement production by statistical operators. The entangling properties of 
the latter define the characteristic features of quantum measurements, as well as
the structure of probability measure in quantum information processing and quantum 
decision theory. Also, it can be necessary to study {\it thermal entanglement 
production} characterized by the amount of entanglement produced by connecting 
an initially closed nonentangled quantum system to a thermal bath. There exists 
a variety of finite quantum systems that can be initially prepared in a desired pure 
state \cite{Birman_33}. Then this system can be connected to a thermal bath in the 
standard sense of realizing a thermal contact that transfers heat but does not 
destroy the system itself, as a result of which the system acquires the thermal Gibbs 
distribution \cite{Kubo_34}. The immediate question is how much entanglement can 
be produced by this nonunitary procedure of connecting an initially closed quantum 
system to a thermal bath? 

Statistical operators of pure states are determined by system wavefunctions. According 
to the Pauli principle \cite{kaplan2013}, many-body wavefunctions of indistinguishable 
particles can be either permutation-symmetric for bosons or antisymmetric for fermions. 
However, additional possibilities appear for spinor particles, which have spin and spatial
degrees of freedom. The spin and spatial wavefunctions can belong to multidimensional, 
non-Abelian, irreducible representations of the symmetric group \cite{hamermesh}, being
combined to the symmetric or antisymmetric total wavefunction \cite{Kaplan,pauncz_symmetric}.  
The non-Abelian permutation symmetry has been considered in the early years of quantum 
mechanics \cite{wigner1927,heitler1927,dirac1929} and applied later in spin-free 
quantum chemistry \cite{Kaplan,pauncz_symmetric}, as well as in other fields 
\cite{lieb1962,yang1967,sutherland1968,guan2009,yang2009,gorshkov2010,fang2011,
daily2012,harshman2014,yurovsky14,harshman2016,harshman2016a,yurovsky2016,
brechet2016,sela2016,yurovsky2017}. 

It is the aim of the present paper to study the entanglement production by statistical
operators. In Sec. 2, we explain the {\it operational meaning of entanglement production}, 
introduce basic notations, concretize the difference between separable and nonentangling 
operators, and  demonstrate how the problem of entanglement production by statistical operators 
naturally arises in the theory of quantum measurements, quantum information processing, and 
quantum decision theory. In Sec. 3, we define a general measure of entanglement production 
by arbitrary operators and specify the consideration for different types of statistical operators. 
The calculational procedure for this measure is demonstrated in Sec. 4 by several simple, but 
important, examples of entangled pure states. We explain in Sec. 5 how the introduced measure 
of entanglement production is connected with the other known quantities, such as the purity of 
quantum states, linear entropy, or impurity, inverse participation ratio, quadratic R\'{e}nyi entropy, 
and the correlation function of composite measurements. This measure can be defined for a set 
of statistical operators characterizing a system after quantum measurements. Section 6 
demonstrates that the decoherence phenomenon is connected with the increase of  the 
entanglement-production measure. In Sec. 7, we study the entanglement production by 
an equilibrium Gibbs operator with the Ising type Hamiltonian, since such Hamiltonians are 
widely employed for representing qubit registers. The measure of entanglement production 
depends on the type of coupling between qubits, whether it is ferromagnetic or antiferromagnetic. 
In Sec. 8, we turn to complex multiparticle systems, for which it is admissible to consider 
different ways of partitioning the system degrees of freedom. In Sec. 9, we study a general 
case of a multiparticle spinor system, calculating the entanglement production measure for 
particle partitioning and for spin-spatial partitioning. Section 10 concludes.

\section{Operational meaning of entanglement production}

In order to avoid confusion, let us first of all concretize the difference between 
separable and nonentangling operators. We also stress the importance of the operator 
entanglement production in the process of quantum measurements \cite{Neumann_58}. 
Note that quantum measurements can be treated as decisions in decision theory 
\cite{Neumann_58,Benioff_59,Holevo_60}, because of which the mathematical structure 
of quantum decision theory is the same as that of quantum measurement theory 
\cite{Yukalov_10,Yukalov_61,Yukalov_62}. The difference is only in terminology, where 
a measurement is called a decision and the result of a measurement is termed an event.

An operator is defined on a Hilbert space and acts on wave functions (vectors) of this space. 
The property of the operator to produce entangled wave functions from disentangled ones is
called entanglement production. {\it The operational meaning of the entangling power of an 
operator is its ability of entangling the wave functions of the Hilbert space it acts on}
\cite{Marcus_13,Westwick_14,Johnston_15,Beasley_16,Alfsen_17,Friedland_18,Gohberg_19,
Crouzeux_20,Chen_23}. This notion is applicable to any operator acting on a Hilbert space, 
including statistical operators. 

\subsection{Separable versus nonentangling operators}

One considers a system in a Hilbert space $\mathcal{H}$ characterized by a statistical
operator $\hat{\rho}$ that is a semi-positive, trace-one operator. The pair
$\{\mathcal{H}, \hat{\rho}\}$ is called {\it statistical ensemble}. The considered system 
is composite, with the Hilbert space being a tensor product 
\be
\label{1}
 \cH = \bigotimes_{i=1}^N \cH_i \;  .
\ee
Each space $\mathcal{H}_i$ possesses a basis $\{\vert n_i \rangle \}$, so that
\be
\label{A4}
  \cH_i = {\rm span}\{ | n_i \rgl \} \; , \qquad
\cH =  {\rm span} \left \{ \bigotimes_{i=1}^N \; | n_i \rgl \right \} \;   .
\ee

An operator algebra $\{ \hat{A} \}$ is defined on the space $\mathcal{H}$, consisting 
of trace-class operators, for which
\be
\label{2}
 0 \neq | {\rm Tr}_\cH \hat A | < \infty \;  .
\ee
An operator $\hat{A}$, acting on a disentangled function of $\mathcal{H}$ can either 
result in another disentangled function or transform the disentangled function into 
an entangled function. The sole type of a nonentangling operator, except the trivial
swap operator changing the labelling, has the factor form   
\cite{Westwick_14,Johnston_15,Friedland_18}
\be
\label{3}
 \hat A_\otimes = \bigotimes_{i=1}^N  \hat A_i \;  ,
\ee
which, as is evident, is defined up to a multiplication constant. 

The notion of separable states can be extended to operators
\cite{Yukalov_6,Zanardi_7,Balakrishnan_8,Macchiavello_9,Yukalov_10,Kong_11}.
Then a separable operator is such that can be represented as the finite linear 
combination
\be
\label{4}
 \hat A_{sep} = \sum_k \lbd_k \bigotimes_{i=1}^N \hat A_{ik} \;  ,
\ee
where $\lambda_k$ are complex-valued numbers. A nonentangling operator is 
a particular case of a separable operator, when $\lambda_k$ is proportional to $\delta_{kk_o}$,
i.e., it is a rank-1 separable operator. But the principal difference of a general separable 
operator from a nonentangling operator is that the former does entangle disentangled functions. 
This is evident from the action of a separable operator on a disentangled function, 
yielding
\be
\label{5}
 \hat A_{sep} \vp_{dis} = \sum_k \lbd_k \bigotimes_{i=1}^N \hat A_{ik} \vp_i \; ,
\ee
which is an entangled function, if $\lambda_k$ is not proportional to $\delta_{kk_o}$.

Observable quantities are represented by self-adjoint operators $\hat{A}$. For 
a system characterized by a statistical operator $\hat{\rho}$, the measurable quantities
are given by the averages
\be
\label{6}
  \lgl \hat A \rgl \equiv {\rm Tr}_\cH \hat\rho \hat A \; .
\ee

The peculiarity of measurements are essentially different for the systems with an 
entangling or nonentangling statistical operators. Even if one is measuring an 
observable corresponding to a nonentangling operator (\ref{3}), but the statistical 
operator being entangling, the related average is not reducible to a product
of partial averages,
\be
\label{7}
 \lgl \hat A_\otimes \rgl \neq \prod_{i=1}^N \lgl \hat A_i \rgl \qquad
(\hat\rho \neq \hat \rho_\otimes ) \;  ,
\ee
where
\be
\label{A5}
 \lgl \hat A_\otimes \rgl = {\rm Tr}_\cH \hat\rho \hat A_\otimes \; , \qquad
 \lgl \hat A_i \rgl = {\rm Tr}_{\cH_i} \hat\rho_i \hat A_i \; .
\ee
Such a reduction is possible only when the statistical operator is also nonentangling.

\subsection{Structure of probability measure}

Similarly, in quantum decision theory, an event is represented by an operator $\hat{P}$
that is either a projector or, more generally, an element of a positive operator-valued
measure \cite{Williams_1,Nielsen_2,Vedral_3,Keyl_4,Wilde_5,Yukalov_6}. The event
operator $\hat{P}$ plays the role of an operator of observable. And the probability of 
the event is defined by the average
\be
\label{8}
  p(\hat P) \equiv \lgl \hat P \rgl = {\rm Tr}_\cH \hat\rho \hat P \; ,
\ee 
which takes the values in the interval $0 \leq p(\hat P) \leq 1$. A composite event, 
describing the set of independent partial events, has the form of a nonentangling operator
\be
\label{9}
  \hat P_\otimes = \bigotimes_{i=1}^N \hat P_i \; .
\ee
If the system statistical operator is entangling, the probability of the composite event
cannot be reduced to the product of the probabilities of partial events,
\be
\label{10}
  p(\hat P_\otimes) \neq \prod_{i=1}^N p(P_i) \qquad 
(\hat\rho \neq \hat\rho_\otimes) \; ,
\ee
where
\be
\label{A6}
p(\hat P_\otimes) = \lgl \hat P_\otimes \rgl \; , \qquad   
p(\hat P_i) = \lgl \hat P_i \rgl \;  .
\ee
The reduction is possible only if the statistical operator is also nonentangling. Thus
the structure of the probability measure is principally different for the cases of either
entangling or nonentangling statistical operators.

\section{Measure of entanglement production}

One usually considers the entangling properties of unitary operators describing gates
acting on bipartite systems, but in general, the operator does not need to be unitary.  
If an operator $\hat{U}$ acts on a bipartite state function $|\varphi_{12} \rangle$, one 
gets a new function  $\hat{U}|\varphi_{12}\rangle$ defining the corresponding bipartite state 
$\hat{\rho}_{12} = \hat{U}|\varphi_{12} \rangle \langle \varphi_{12}|\hat{U}^+$. Then the
problem is reduced to studying the entangled structure of this state $\hat{\rho}_{12}$ 
by means of the known entanglement measures of bipartite states, such as entangling 
power, linear entropy, and like that \cite{Zanardi_7,Macchiavello_9,Kong_11,Chen_63}.
However, this does not describe the global entangling property of an operator on
the whole Hilbert space where it is defined. 

A general measure of entanglement production, applicable to arbitrary
(not necessarily unitary) operators acting on the whole Hilbert space, containing any
number of factors, has been suggested in Refs. \cite{Yukalov_24,Yukalov_25}. Here,
we shall use this measure for quantifying the entangling properties of statistical
operators.

\subsection{Arbitrary operators}

The definition of the measure is as follows. Let us be interested in the entangling
properties of an operator $\hat{A}$ acting on a composite Hilbert space (\ref{1}).
The idea is to compare the action of this operator on $\mathcal{H}$ with the action of
its nonentangling product counterpart
\be
\label{11}
\hat A_\otimes \equiv C \bigotimes_{i=1}^N \hat A_i
\ee
that is a product of the reduced operators
\be
\label{12}
 \hat A_i \equiv {\rm Tr}_{\cH/\cH_i} \hat A \;  ,
\ee
where the trace is over all $\mathcal{H}$ except the subspace $\mathcal{H}_i$. The
constant $C$ is defined by the normalization condition
\be
\label{13}
{\rm Tr}_\cH \hat A_\otimes = {\rm Tr}_\cH \hat A  ,
\ee
which gives $C = \left ( {\rm Tr}_\cH \hat A \right )^{1-N}$. Therefore
\be
\label{14}
\hat A_\otimes = \frac{\bigotimes_{i=1}^N \hat A_i}{\left ( {\rm Tr}_\cH \hat A \right )^{N-1}} \; .
\ee
By the theorem proved for binary products \cite{Marcus_13,Beasley_16,Alfsen_17},
as well as for an arbitrary number of factors \cite{Westwick_14,Johnston_15,Friedland_18},
the product operator form (\ref{14}) never entangles any functions.

The entanglement production measure for the operator $\hat{A}$ is defined as
\be
\label{15}
 \ep(\hat A) \equiv \log \; \frac{||\hat A||}{||\hat A_\otimes|| } \;  .
\ee
The logarithm can be taken with respect to any base. This quantity (\ref{15}) satisfies 
all conditions required for being classified as a measure \cite{Yukalov_24,Yukalov_25,Yukalov_79}.
Thus it enjoys the properties: (i) it is semipositive and bounded for the finite number of factors $N$; 
(ii) it is continuous in the sense of norm convergence; (iii) it is zero for nonentangling operators;
(iv) it is additive; (v) it is invariant under local unitary operations.   
 
As the norm here, it is convenient to accept the Hilbert-Schmidt norm
\be
\label{16}
 ||\hat A|| = \sqrt{ {\rm Tr}_\cH (\hat A^+ \hat A)   } \;  ,
\ee
which does not depend on the chosen basis. Respectively, the norm of a partial
reduced operator, acting on $\mathcal{H}_i$, is
\be
\label{17}
 ||\hat A_i || = \sqrt{ {\rm Tr}_{\cH_i} (\hat A_i^+ \hat A_i)   } \;  .
\ee

\subsection{Statistical operators}

Our aim is to consider statistical operators, for which the nonentangling counterpart
reads as
\be
\label{18}
\hat{\rho}_\otimes = \bigotimes_{i=1}^N \hat{\rho}_i \; , \qquad  \hat{\rho}_i \equiv  {\rm Tr}_{\cH/\cH_i} \hat{\rho} \; .
\ee
The normalization condition is
\be
\label{19}
 {\rm Tr}_\cH  \hat\rho_\otimes = {\rm Tr}_\cH \hat\rho = 1 \;  .
\ee
Therefore we need to study the measure
\be
\label{20}
\ep(\hat\rho) = \log \; \frac{||\hat\rho||}{||\hat\rho_\otimes||} \; ,
\ee
in which
\be
\label{3.11}
 ||\hat\rho_\otimes|| = \prod_{i=1}^N ||\hat\rho_i||  = 
\prod_{i=1}^N \sqrt{ {\rm Tr}_{\cH_i} \hat\rho_i^2}  \; .
\ee
Explicitly, the measure writes as
\be
\label{3.12}
\ep(\hat\rho) = \frac{1}{2}\; \log\; 
\frac{{\rm Tr}_\cH\hat\rho^2}{\prod_{i=1}^N{\rm Tr}_{\cH_i}\hat\rho_i^2} \;   .
\ee

\subsection{Pure states}

In the case of pure states, statistical operators have the form
\be
\label{21}
  \rho = \vert \psi \rangle \langle \psi \vert  \; ,             
\ee
where $\vert \psi \rangle$ is a normalized wave function. This statistical operator 
is idempotent, so that
\be
\label{A8}
 ||\hat\rho || = \sqrt{ {\rm Tr}_\cH \hat\rho^2} = 1 \qquad
(\hat\rho^2 = \hat\rho ) \;  .
\ee
Then measure (\ref{20}) becomes
\be
\label{22}
 \ep(\hat\rho) = -\log || \hat\rho_\otimes || \;  .
\ee
Or, taking into account the above relations, for pure states, we get
\be
\label{23}
 \ep(\hat\rho) = - \sum_{i=1}^N  \log || \hat\rho_i || = - \;
\frac{1}{2} \sum_{i=1}^N \log {\rm Tr}_{\cH_i} \hat\rho_i^2 \; .
\ee

Notice that the defined measure is valid for arbitrary systems, with any statistical
operators, and with any number of factors in the Hilbert space (\ref{1}). Also, any
operators, unitary or not, can be considered \cite{Yukalov_25}.

\subsection{Separable states}

As has been mentioned in Sec. II, separable operators are, generally, entangling.
Now, we can demonstrate this by explicitly calculating the measure of 
entanglement production for a separable statistical operator. Let us consider a 
separable state
\be
\label{24}
 \hat\rho_{sep} = \sum_k p_k \bigotimes_{i=1}^N \hat\rho_{ik} \qquad
(\hat\rho_{ik} = | n_{ik} \rgl \lgl n_{ik} | ) \;  ,
\ee
where normalized wave functions $| n_{ik} \rgl$ belong to ${\cH_i} $ and are orthogonal, 
$\lgl n_{i k'} | n_{ik} \rgl =\delta_{kk'}$. Taking into account the properties
\be
\label{A13}
 \hat\rho_{ik} \hat\rho_{ip} =\dlt_{kp} \hat\rho_{ik} \; , \qquad 
{\rm Tr}_{\cH_i} \hat\rho_{ik} = 1 \; ,  \qquad 
\hat\rho_{sep}^2 = \sum_k p_k^2 \bigotimes_{i=1}^N \hat\rho_{ik} \; ,
\ee
we get the norm 
\be
\label{A14}
 ||  \hat\rho_{sep} || = \sqrt{\sum_k p_k^2 } \; .
\ee
The partial statistical operators are
\be
\label{A15}
\hat\rho_i \equiv {\rm Tr}_{\cH/\cH_i} \hat\rho_{sep} = \sum_k p_k \hat\rho_{ik} \;  ,
\ee
with the properties
\be
\label{A16}
\hat\rho_i^2 =  \sum_k p_k^2 \hat\rho_{ik} \; , \qquad 
|| \hat\rho_i || =  \sqrt{\sum_k p_k^2 } \; .
\ee
Then for the norm of the nonentangling counterpart, we find
\be
\label{A17}
 || \hat\rho_\otimes || = \left ( \sum_k p_k^2 \right )^{N/2} \; .
\ee
The entanglement production measure (\ref{20}) becomes
\be
\label{25}
 \ep(\hat\rho_{sep} ) = -\; \frac{N-1}{2} \; \log  \sum_k p_k^2 \;  .
\ee
This is evidently nonzero, provided that $N > 1$ and $p_k \neq \delta_{kk_0}$.

\subsection{Gibbs states}

For an equilibrium system, characterized by a Hamiltonian $H$, the Gibbs statistical
operator is
\be
\label{26}
 \hat\rho = \frac{1}{Z} \; e^{-\bt H} \;  ,  Z = {\rm Tr}_\cH e^{-\bt H} \; ,
\ee
where $\beta$ is inverse temperature. With the partial operators
\be
\label{A18}
\hat\rho_i = \frac{1}{Z} \; {\rm Tr}_{\cH/\cH_i} e^{-\bt H} \; ,
\ee
the nonentangling counterpart is
\be
\label{27}
 \hat\rho_\otimes =  \frac{1}{Z^N} \; \bigotimes_{i=1}^N {\rm Tr}_{\cH/\cH_i} e^{-\bt H} \; .
\ee
Introducing the notations
\be
\label{28}
f_1 \equiv || e^{-\bt H} ||^2 = {\rm Tr}_{\cH} e^{-2\bt H} 
\ee
and 
\be
\label{29}
 f_2 \equiv 
\prod_{i=1}^N {\rm Tr}_{\cH_i} \left ( {\rm Tr}_{\cH/\cH_i} e^{-\bt H} \right )^2 \; ,
\ee
we can represent the entanglement production measure as
\be
\label{30}
 \ep(\hat\rho) = \frac{1}{2} \; \log \left ( \frac{f_1}{f_2} \; Z^{2N-2} \right ) \; .
\ee
Thus, for a given Hamiltonian, we need to calculate the functions (\ref{28}) and (\ref{29}),
and the partition function $Z$.

\section{Entangled pure states}

Before going to more complicated problems, it is useful to illustrate how the measure 
is calculated for simple cases of pure states. Generally, depending on the definition
of the employed norm, the entanglement production measure can be slightly different
\cite{Yukalov_25}. Here we use the Hilbert-Schmidt norm. We shall see that for 
bipartite systems with entangled states, the entanglement production measure
$\varepsilon(\hat{\rho})$ coincides with the entanglement von Neumann entropy 
$S(\hat{\rho}_i) \equiv - \rm{Tr}_{\cH_i}  \hat{\rho}_i \ln \hat{\rho}_i$. The examples 
considered in this section illustrate how the measure is calculated, which will allow 
us to shorten the explanation of intermediate calculations in the following more 
complicated cases.

\subsection{Einstein-Podolsky-Rosen states}

The corresponding statistical operator is
\be
\label{31}
 \hat\rho_{EPR} = | EPR \rgl \lgl EPR | \;  ,
\ee
where
\be
\label{A19}
 | EPR \rgl = \frac{1}{\sqrt{2}} \; ( | 12 \rgl \pm |21 \rgl ) \;  .
\ee
The reduced operators are
\be
\label{A20}
\hat\rho_i = \frac{1}{2} \; ( | 1 \rgl \lgl 1 | + | 2 \rgl \lgl 2 | ) \;   ,
\ee
for which
\be
\label{A21}
 \hat\rho_i^2 = \frac{1}{4} \; ( | 1 \rgl \lgl 1 | + | 2 \rgl \lgl 2 | ) \;  .
\ee
The corresponding norms are
\be
\label{A22}
|| \hat\rho_i || = \frac{1}{\sqrt{2}} \; , \qquad 
|| \hat\rho_\otimes || = \frac{1}{2} \; .
\ee
Then we find the entanglement production measure
\be
\label{32}
 \ep(\hat\rho_{EPR}) = \log 2 \;  .
\ee

Note that in this case, the entanglement entropy
$S(\hat{\rho}_i) \equiv - \rm{Tr}_{\cH_i}  \hat{\rho}_i \ln \hat{\rho}_i$ coincides
with measure (\ref{20}).

\subsection{Bell states}

The statistical operator is
\be
\label{33}
 \hat\rho_B = | B \rgl \lgl B | \;  ,
\ee
where
\be
\label{A23}
 | B \rgl = \frac{1}{\sqrt{2}} \; ( | 11 \rgl \pm | 22 \rgl ) \;  .
\ee
Calculations are similar to the previous case, giving
\be
\label{34}
 \ep(\hat\rho_B ) = \log 2 \;  .
\ee
Again, this coincides with the entanglement entropy $S(\hat{\rho}_i) = \log 2$.

\subsection{Greenberger-Horne-Zeilinger states}

These states are a generalization of two-particle Bell states to $N$ particles, so that
\be
\label{35}
 \hat\rho_{GHZ} = | GHZ \rgl \lgl GHZ | \;  ,
\ee
with
\be
\label{A24}
 | GHZ \rgl  = \frac{1}{\sqrt{2}} \; ( | 11\ldots 1 \rgl \pm | 22\ldots 2 \rgl ) \; .
\ee
Now we have
\be
\label{A25}
 \hat\rho_\otimes = \bigotimes_{i=1}^N \hat\rho_i \; , \qquad
|| \hat\rho_i || = \frac{1}{\sqrt{2}} \; , \qquad
|| \hat\rho_\otimes || = \prod_{i=1}^N || \hat\rho_i || = \frac{1}{2^{N/2}} \;  .
\ee
As a result
\be
\label{36}
 \ep(\hat\rho_{GHZ} ) = \frac{N}{2} \; \log 2 \qquad ( N \geq 2 ) \;  .
\ee
The entanglement entropy for $N$-particle states is not defined.

\subsection{Multicat states}

Such states are a generalization of the Schr\"{o}dinger cat states to $N$ objects,
\be
\label{37}
 \hat\rho_{MC} = | MC \rgl \lgl MC | \;  ,
\ee
where
\be
\label{A26}
 | MC \rgl = c_1 | 11 \ldots 1 \rgl + c_2 | 22\ldots 2 \rgl \;  ,
\ee
with $c_i$ being complex numbers satisfying the normalization $| c_1|^2 + |c_2|^2 = 1$ .
The reduced operators are
\be
\label{A27}
 \hat\rho_i = | c_1|^2 | 1 \rgl \lgl 1 | +  | c_2 |^2 | 2 \rgl \lgl 2 | \; .
\ee
Calculating the norms
\be
\label{A28}
 || \hat\rho_i || = \sqrt{ {\rm Tr}_{\cH_i} \hat\rho_i^2 } = 
\sqrt{ |c_1|^4 + |c_2|^4 } \;  , \qquad 
 || \hat\rho_\otimes || = \prod_{i=1}^N  || \hat\rho_i || = 
\left (  |c_1|^4 + |c_2|^4 \right )^{N/2} \; ,
\ee
we obtain
\be
\label{38}
 \ep(\hat\rho_{MC} ) = -\; \frac{N}{2} \; \log \left (  |c_1|^4 + |c_2|^4 \right ) \;  .
\ee
The maximal entanglement production is reached for $|c_i|^2 = 1/2$, yielding
\be
\label{A29}
\sup \ep(\hat\rho_{MC}) = \frac{N}{2} \; \log 2 \qquad ( N \geq 2 ) \; .
\ee

\subsection{Multimode states}

These states are a generalization of the multicat states, when each object can be
not in two, but in $M$ different modes,
\be
\label{39}
 \hat\rho_{MM} = | MM \rgl \lgl MM | \;  ,
\ee
where
\be
\label{A30}
| MM \rgl  =\sum_{n=1}^M c_n | nn \ldots n \rgl \;   ,
\ee
with the coefficients satisfying the normalization
\be
\label{A31}
 \sum_{n=1}^M | c_n |^2 = 1 \;  .
\ee
The reduced operators become
\be
\label{A32}
 \hat\rho_i  =   \sum_{n=1}^M | c_n |^2 | n \rgl \lgl n | \; .
\ee
With the norms
\be
\label{A33}
 || \hat\rho_i   || = \sqrt{ \sum_{n=1}^M | c_n |^4 } \; , \qquad
 || \hat\rho_\otimes || = \left ( \sum_{n=1}^M | c_n |^4 \right )^{N/2} \; ,
\ee
we derive
\be
\label{40}
 \ep(\hat\rho_{MM} ) = - \; \frac{N}{2} \; \log \sum_{n=1}^M | c_n |^4 \;  .
\ee
The maximal entanglement production happens for $|c_i|^2 = 1/M$, resulting in
\be
\label{A34}
\sup  \ep(\hat\rho_{MM} ) = \frac{N}{2} \; \log M \;  .
\ee

\section{Relation to other concepts}

The meaning of measure(\ref{3.12}) can be better understood by studying its connection with 
other important quantities employed in quantum theory 
\cite{Williams_1,Nielsen_2,Vedral_3,Keyl_4,Horodecki_76,Guhne_77,Wilde_5,Siewert_78,Yukalov_6}.  
Below we show these connections with the most often met concepts. 

\subsection{Purity of quantum state}

The purity of a quantum state $\hat{\rho}$ in the Hilbert space $\mathcal{H}$ is defined as
\be
\label{5.1}
 \gm(\hat\rho) \equiv {\rm Tr}_\cH \hat\rho^2 \;  .
\ee
It varies in the interval
\be
\label{5.2}
  \frac{1}{d} \leq \gm(\hat\rho) \leq 1 \qquad ( d \equiv {\rm dim}\cH ) 
\ee
and shows the closeness of the state to a pure state. For a pure state $\gm(\hat\rho)=1$,
while for a completely mixed state $\gamma(\hat\rho) = 1/d$. Purity describes the spread 
of the state over the given basis.
 
Similarly to the purity of the total state $\hat{\rho}$, it is possible to introduce the 
purity of the partial states
\be
\label{5.3}
\gm(\hat\rho_i) \equiv {\rm Tr}_{\cH_i} \hat\rho^2_i 
\ee
varying in the interval
\be
\label{5.4}
 \frac{1}{d_i} \leq \gm(\hat\rho_i) \leq 1 \qquad ( d_i \equiv {\rm dim}{\cH_i} )\; .
\ee
The purity of the nonentangling state $\hat{\rho}_\otimes$ becomes
\be
\label{5.5}
 \gm(\hat\rho_\otimes) \equiv {\rm Tr}_\cH \hat\rho_\otimes^2 = 
\prod_{i=1}^N \gm(\hat\rho_i) \; .
\ee

Then measure (\ref{3.12}) can be presented as
\be
\label{5.6}
 \ep(\hat\rho) = \frac{1}{2}\; \log \; \frac{ \gm(\hat\rho)}{\gm(\hat\rho_\otimes)}\; ,
\ee
varying in the interval
\be
\label{5.7}
0 \leq  \ep(\hat\rho) \leq \frac{1}{2} \; \log d \qquad 
\left ( d = \prod_{i=1}^N d_i \right ) \;  .
\ee

The denominator of the fraction under the logarithm in equation (\ref{5.6}) has the
meaning of an effective purity of the nonentangling state of a system composed of 
partial subsystems. Hence measure  (\ref{5.6}) shows how much the purity of the given 
state $\hat{\rho}$ is larger than the effective purity of the nonentangling state
$\hat{\rho}_\otimes$ corresponding to the system composed of partial subsystems.

\subsection{Linear entropy or impurity}

The linear entropy of a state $\hat{\rho}$ is given by the expression
\be
\label{5.8}
  S_L(\hat\rho) \equiv 1 - {\rm Tr}_\cH \hat\rho^2 = 1 - \gm(\hat\rho) \; ,
\ee
which varies in the interval
\be
\label{5.9}
  0 \leq S_L(\hat\rho) \leq 1 \; - \; \frac{1}{d} \; .
\ee
Because of its relation (\ref{5.8}) to the state purity, the linear entropy is also 
called impurity. In the same way, the linear entropy of the nonentangling state 
$\hat{\rho}_\otimes$ is
\be
\label{5.10}
 S(\hat\rho_\otimes) = 1 - {\rm Tr}_\cH \hat\rho^2_\otimes = 
1 - \gm(\hat\rho_\otimes) \;  .
\ee
Therefore, measure (\ref{3.12}) can be written as
\be
\label{5.11}
 \ep(\hat\rho) = 
\frac{1}{2} \; \log \; \frac{1-S_L(\hat\rho)}{1-S_L(\hat\rho_\otimes)} \;  .
\ee
By partitioning the system, with a state $\hat{\rho}$, into subsystems, with 
partial states $\hat{\rho}_i$, one gets the system composed of the subsystems, with 
the nonentangling state $\hat{\rho}_\otimes$, whose purity is smaller than the purity 
of the initial state $\hat{\rho}$. Consequently, the impurity, that is, the linear 
entropy, of the nonentangling state $\hat{\rho}_\otimes$ is larger than that of the 
state $\hat{\rho}$. In that sense, measure (\ref{5.11}) describes how much the impurity 
of the state $\hat{\rho}_\otimes$ increases, as compared to the state $\hat{\rho}$, 
before the partitioning.

\subsection{Inverse participation ratio}

Sometimes purity is used as a measure of localization and linear entropy, as a measure
of delocalization. This is because these concepts are closely connected with the notion
of inverse participation ratio \cite{Dean_84,Edwards_85,Heller_86,Yurovsky_11,Olshanii_12}. 

Inverse participation ratio can be introduced as a measure of localization in the 
real space for characterizing Anderson localization or in phase space for describing 
semiclassical localization. For the purpose of the present paper, it is more convenient 
to introduce the inverse participation ratio characterizing Hilbert-space localization 
\cite{Cohen_87}, which can be defined as
\be
\label{5.12}
 R^{-1}(\hat\rho) \equiv \lim_{\tau\ra\infty} \; \frac{1}{\tau} \int_0^\tau
{\rm Tr}_\cH \hat\rho(0)\hat\rho(t) \; dt \;  ,
\ee
where
$$
 \hat\rho(t) = e^{-i\hat H t}\; \hat\rho(0)\; e^{i\hat H t} \;  .
$$
This definition shows that the inverse participation ratio is equivalent to the transition
probability averaged over time. 

For the basis formed by the eigenvectors of the Hamiltonian from the eigenproblem
$$
 \hat H \; | \; n \; \rgl =  E_n \; | \; n \; \rgl \; ,
$$
we find 
\be
\label{5.13}
  R^{-1}(\hat\rho) = \lim_{\ep\ra 0} \; \sum_{mn} \; 
\frac{\ep^2\rho_{mn}\rho_{nm}}{\ep^2+(E_n-E_m)^2} \; ,
\ee
in which $\rho_{mn} \equiv \langle m | \hat{\rho} (0) | n \rangle$. This expression is 
valid for both nondegenerate or degenerate spectrum. For a nondegenerate spectrum, this 
simplifies to
\be
\label{5.14}
R^{-1}(\hat\rho) = \sum_n \rho_{nn}^2 \;   .
\ee
The inverse participation ratio varies in the range
\be
\label{5.15}
 \frac{1}{d} \leq R^{-1}(\hat\rho) \leq 1 \;  ,
\ee
which is the same as the variation range of state purity. Remembering definition (\ref{5.1}) 
of state purity and introducing the notation for the second-order coherence function \cite{Baumgratz_91}
$$
 C_2(\hat\rho) \equiv \sum_{m\neq n} | \rho_{mn} |^2   \;  ,
$$
we obtain the relation
\be
\label{5.16}      
R^{-1}(\hat\rho) = \gm(\hat\rho) - C_2(\hat\rho)
\ee
between the purity and inverse participation ratio. When the nondiagonal terms are less
important than the diagonal ones, the inverse participation ratio is approximately equal
to the state purity. This is why the latter can also serve as a measure characterizing 
localization in a Hilbert space. Therefore measure (\ref{3.12}) shows to what extent
the system state $\hat{\rho}$ is more localized in the Hilbert space than the state 
$\hat{\rho}_\otimes$ of the partitioned system.

\subsection{Quantum R\'{e}nyi entropy}

In the quantum setting, the R\'{e}nyi entropy of order $\alpha$, for a state $\hat{\rho}$
acting on a Hilbert space $\mathcal{H}$, is given by the form
\be
\label{5.17}
 H_\al(\hat\rho) \equiv \frac{1}{1-\al} \; \log\; {\rm Tr}_\cH \hat\rho^\al \; .
\ee
Here we are interested in the quadratic R\'{e}nyi entropy
\be
\label{5.18}
 H_2(\hat\rho)  = - \log\; {\rm Tr}_\cH \hat\rho^2 = - \log \gm(\hat\rho) \; ,
\ee
which is connected with the state purity and varies in the range
\be
\label{5.19}
  0 \leq  H_2(\hat\rho) \leq \log d \; .
\ee
Since the state purity can characterize Hilbert-space localization, the quadratic 
R\'{e}nyi  entropy can serve as a measure of impurity and delocalization 
\cite{Mirbach_88,Muller_80,Calixto_89}.  

The quadratic R\'{e}nyi entropy can also be defined for partial states,
\be
\label{5.20}
H_2(\hat\rho_i)  = - \log\; {\rm Tr}_{\cH_i} \hat\rho^2_i = - \log \gm(\hat\rho_i) \;   ,
\ee
being in the range
\be
\label{5.21}
  0 \leq  H_2(\hat\rho_i) \leq \log d_i \; .
\ee
Then the quadratic R\'{e}nyi entropy for the partitioned state $\hat{\rho}_\otimes$
reads as
\be
\label{5.22}
 H_2(\hat\rho_\otimes)  = - \log\; {\rm Tr}_\cH \hat\rho^2_\otimes = 
\sum_{i=1}^N H_2(\hat\rho_i) \;   .
\ee
Therefore measure (\ref{3.12}) can be represented as the difference
\be
\label{5.23}
 \ep(\hat\rho) = \frac{1}{2} \; 
\left [ H_2(\hat\rho_\otimes) - H_2(\hat\rho) \right ] \;  ,
\ee
quantifying how much the R\'{e}nyi entropy of the partitioned state $\hat{\rho}_\otimes$
is larger than that of the initial state $\hat{\rho}$. In other words, measure (\ref{5.23})
is half of the difference, measured in terms of the R\'{e}nyi entropy, of the state $\hat{\rho}$
from the product state $\hat{\rho}_\otimes$. 

For a pure state $\hat{\rho}$, the R\'{e}nyi entropy is zero,
\be
\label{5.24}
 H_2(\hat\rho) = 0 \qquad ( \hat\rho^2 = \hat\rho ) \; .
\ee
Then for a pure state, measure (\ref{5.23}) becomes one half of the sum of partial 
R\'{e}nyi entropies
\be
\label{5.25}
 \ep(\hat\rho) = \frac{1}{2} \; H_2(\hat\rho_\otimes) = 
\frac{1}{2} \sum_{i=1}^N H_2(\hat\rho_i) \qquad ( \hat\rho^2 = \hat\rho ) \; .
\ee

In the case of a bipartite system, the partial R\'{e}nyi entropies
$$
H_2(\hat\rho_1) = H_2(\hat\rho_2) \qquad ( N = 2 )
$$
play the role of the system entanglement entropies. In such a case, measure (\ref{5.25})
coincides with the entanglement entropy,
\be
\label{5.26}
 \ep(\hat\rho) = H_2(\hat\rho_i) \qquad ( \hat\rho^2 = \hat\rho\;, ~ N = 2 ) \;  .
\ee
Notice that for bipartite systems the R\'{e}nyi entropy is claimed to be available for 
measuring \cite{Linke_90}.

Recall that in the general case, measure (\ref{5.23}) quantifies how much the R\'{e}nyi 
entropy of the partitioned state $\hat{\rho}_\otimes$ overweights the R\'{e}nyi entropy 
of the initial state $\hat{\rho}$. Since the R\'{e}nyi entropy shows the degree of 
delocalization, the measure (\ref{5.23}) defines to what extent the partitioned state 
$\hat{\rho}_\otimes$ is more delocalized than the initial state $\hat{\rho}$.

\subsection{Correlation in composite measurements}

Entangling property of a statistical operator is of great importance for studying 
correlations in composite measurements. For simplicity, we consider here a bipartite 
system, with the Hilbert space
\be
\label{5.27}
 \cH = \cH_A \bigotimes \cH_B \;  ,
\ee
although the generalization to larger composite systems is straightforward. 

Let us examine a composite measurement, represented by the operator 
$\hat{A} \bigotimes \hat{B}$, formed by two measurements described by the operators
$\hat{A}$ on $\mathcal{H}_A$ and $\hat{B}$ on $\mathcal{H}_B$, respectively. The 
operators $\hat{A}$ and $\hat{B}$ correspond to the operators of local observables.

The correlation between these two measurements is characterized by the correlation 
function
\be
\label{5.28}
 C_{AB} \equiv \lgl  \hat A \bigotimes \hat B  \rgl - 
\lgl \hat A \rgl \lgl \hat B \rgl \; ,
\ee
which explicitly reads as 
\be
\label{5.29}
 C_{AB} = {\rm Tr}_\cH \; \hat\rho \; \hat A \bigotimes \hat B - 
\left ( {\rm Tr}_{\cH_A} \hat\rho_A \hat A \right )
\left ( {\rm Tr}_{\cH_B} \hat\rho_B \hat B \right )  ,
\ee
where
$$
\hat\rho_A \equiv {\rm Tr}_{\cH_B} \hat\rho \; , \qquad 
\hat\rho_B \equiv {\rm Tr}_{\cH_A} \hat\rho \;   .
$$
 
As is clear, the value of the correlation function depends on the entangling property 
of the system state $\hat{\rho}$. If the system state is nonentangling, such that it 
can be represented as a product of the partial states, then the correlation function 
is zero,
\be
\label{5.30}
 C_{AB} = 0 \qquad (\hat\rho = \hat\rho_A \otimes \hat\rho_B ) \;  .
\ee
But if the system state $\hat{\rho}$ is entangling, the correlation function is not 
zero, which implies that the two measurements cannot be made independently of each 
other, since they are correlated with each other. The stronger the entangling ability 
of $\hat{\rho}$, that is, the larger its entanglement production measure (\ref{20}), 
the larger the absolute value $|C_{AB}|$ of the correlation function (\ref{5.28}).

\subsection{State reduction after measurements}

For each system state $\hat{\rho}$, we can define the measure of entanglement 
production $\varepsilon (\hat{\rho})$.  Moreover, if the system is subject to measurements, 
then there appear the whole set of possible states and, respectively, the set of the 
related measures. 

Let the system be in a state $\hat{\rho}$. And let us be interested in an observable
represented by the operator $\hat{Q}$ acting on the Hilbert space $\mathcal{H}$. The 
basis of this space can be taken as defined by the eigenproblem
\be
\label{5.31}
 \hat Q \; | \; n \; \rgl = Q_n \; | \; n \; \rgl \;  ,
\ee
with $n$ being the multi-index
$$
n = \{ n_i : \; i = 1,2, \ldots, N \} \; , \qquad  
n_i = \{ n_i^\al : \; \al = 1,2, \ldots, d_i \} \;  .
$$

If the result of the measurement of this observable is $Q_n$, then, according to the
von Neumann - L\"{u}ders theory \cite{Neumann_58,Luders_92}, the system state reduces to
\be
\label{5.32}
  \hat\rho_n = \frac{\hat P_n \hat\rho \hat P_n}{{\rm Tr}_\cH(\hat\rho \hat P_n)} \; .
\ee
Generally, the operators $\hat{P}_n$ here are the projectors on subspaces associated with 
the eigenvalues $Q_n$. For a nondegenerate spectrum of $Q_n$, which we assume for 
simplicity in what follows, $\hat{P}_n = |n \rangle \langle n|$. 

For the new system state $\hat{\rho}_n$, we have
$$
 {\rm Tr}_\cH \hat\rho_n^2 = 
\frac{{\rm Tr}_\cH(\hat\rho\hat P_n)^2}{({\rm Tr}_\cH\hat\rho\hat P_n)^2} \;  ,
$$
where
$$
 {\rm Tr}_\cH (\hat\rho\hat P_n)^2 = \rho^2_{nn} \; , \qquad
  {\rm Tr}_\cH \hat\rho\hat P_n = \rho_{nn} \; , \qquad 
\rho_{mn} \equiv \lgl \; m \; | \; \hat\rho\; | \; n \; \rgl \; .
$$
Hence ${\rm Tr}_{\cH} \hat\rho_n^2 = 1$. The corresponding product state is
$$
 \hat\rho_{n\otimes} = \bigotimes_{i=1}^N \hat\rho_{ni} \; , \qquad
\hat\rho_{ni} \equiv  {\rm Tr}_{\cH/\cH_i} \hat\rho_n \;  .
$$       
Thus, the entanglement production measure of the new state is
\be
\label{5.33}
 \ep(\hat\rho_n) = - \; \frac{1}{2} \; \log  {\rm Tr}_\cH \hat\rho_{n\otimes}^2 =
- \; \frac{1}{2} \sum_{i=1}^N \log  {\rm Tr}_{\cH_i} \hat\rho_{ni}^2 \; .
\ee
Altogether, we get a set of the measures for different multi-indices $n$.

\section{Decoherence in nonequilibrium systems}

In nonequilibrium systems, the state $\hat{\rho}(t)$ depends on time, which can lead
to the temporal evolution of the measure $\varepsilon(\hat{\rho}(t))$. This evolution is 
closely connected with such an important phenomenon as {\it decoherence} \cite{Zurek_93}.
Below, we show that the phenomenon of decoherence is in intimate relation to the 
measure $\varepsilon(\hat{\rho}(t))$.

Let us consider a composite system consisting of two parts and characterized by a
statistical operator $\hat{\rho}(t)$ on a Hilbert space 
$\mathcal{H} = \mathcal{H}_A \bigotimes \mathcal{H}_B$, such that
\be
\label{6.1}
 \cH_A = {\rm span} \{ | \; n \; \rgl \} \; , \qquad
 \cH_B = {\rm span} \{ | \; \al \; \rgl \} \;.
\ee
Suppose we are interested in the subsytem with the space $\mathcal{H}_A$, while the
other part describing what is called surrounding. The latter can include measuring devices.  
Self-adjoint operators of observables, say $\hat{A}$, defined on $\mathcal{H}_A$, 
correspond to the observable quantities given by the average
\be
\label{6.2} 
 \lgl \; \hat A(t) \; \rgl = {\rm Tr}_\cH \hat A(t) \hat\rho(0) =
 {\rm Tr}_\cH \hat\rho(t) \hat A(0) \; .
\ee
This yields
\be
\label{6.3}
 \lgl \; \hat A(t) \; \rgl = \sum_{mn} \rho_{mn}(t) A_{nm} \;  ,
\ee
where
\be
\label{6.4}
  \rho_{mn}(t) = \sum_\al \rho_{mn}^{\al\al}(t) \; , \qquad
\rho_{mn}^{\al\bt}(t) \equiv \lgl \; m\al \; |\; \hat\rho(t) \; | \; n \bt \; \rgl \; , \qquad
A_{mn} \equiv  \lgl \; m \; | \hat A(0) \; | \; n \; \rgl \; .
\ee
Generally, the observable quantity (\ref{6.3}) can be written as the sum of a diagonal 
and nondiagonal terms
\be
\label{6.5}
 \lgl \; \hat A(t) \; \rgl = 
\sum_n \rho_{nn}(t) A_{nn} + \sum_{m\neq n} \rho_{mn}(t)A_{nm} \;  .
\ee
The effect of decoherence implies \cite{Zurek_93} that the nondiagonal term tends to zero 
with time, so that 
\be
\label{6.6}
 \lim_{t\ra\infty} \rho_{mn}(t) = 0 \qquad ( m \neq n) \;  .
\ee
This happens because of the interaction between the subsytem of interest and the surrounding.
Decoherence appears even when the surrounding is represented by measuring devices realizing
the so-called nondestructive, nondemolition, or minimally disturbing measurements 
\cite{Braginsky_94,Yukalov_95,Yukalov_96,Yukalov_97}. 

Calculating the measure   
\be
\label{6.7}
 \ep(\hat\rho(t) ) = 
\log \; \frac{||\hat\rho(t)||}{||\hat\rho_A(t)||\; ||\hat\rho_B(t)||} \; ,
\ee
we have
\be
\label{6.8}
 ||\hat\rho(t)||^2 = {\rm Tr}_\cH \hat\rho^2(t) = \sum_{mn} \; \sum_{\al\bt} 
| \rho_{mn}^{\al\bt}(t) |^2 \; .
\ee
For the partial statistical operators
\be
\label{6.9}
 \hat\rho_A(t) \equiv {\rm Tr}_{\cH_B} \hat\rho(t) = 
\sum_\al \lgl \; \al \; | \; \hat\rho(t) \; | \; \al \; \rgl \; , \qquad
\hat\rho_B(t) \equiv {\rm Tr}_{\cH_A} \hat\rho(t) = 
\sum_n \lgl \; n \; | \; \hat\rho(t) \; | \; n \; \rgl \; ,
\ee
we find 
\be
\label{6.10}
||\hat\rho_A(t) ||^2 \equiv {\rm Tr}_{\cH_A} \hat\rho_A^2(t) = 
\sum_{mn} | \rho_{mn}(t) |^2 \; ,
\qquad
||\hat\rho_B(t) ||^2 \equiv {\rm Tr}_{\cH_B} \hat\rho_B^2(t) = 
\sum_{\al\bt} | \rho^{\al\bt}(t) |^2 \;  ,
\ee
where
\be
\label{6.11}
 \rho^{\al\bt}(t) = \sum_n \rho^{\al\bt}_{nn}(t) \; .
\ee 

With the evolution of the whole system given by the law
\be
\label{6.12}
 \hat\rho(t) = \hat U(t) \; \hat\rho(0) \; \hat U^+(t) \; , \qquad 
\hat U = e^{-i\hat H t} \;  ,
\ee
we get
\be
\label{6.13}
 {\rm Tr}_\cH \hat\rho^2(t) = {\rm Tr}_\cH \hat\rho^2(0) \;  .
\ee

We can choose as the basis, the set of the eigenvectors of the system Hamiltonian,
defined by the eigenproblem
\be
\label{6.14}
\hat H \; | \; n \al \; \rgl = E_{n\al} \; | \; n \al \; \rgl \;  .
\ee
Then we obtain the matrix elements
\be
\label{6.15}
 \rho_{mn}^{\al\bt}(t) \equiv \rho_{mn}^{\al\bt}(0) \exp (- i \om_{mn}^{\al\bt} t ) \; ,
\ee
in which
\be
\label{6.16}
  \om_{mn}^{\al\bt} \equiv E_{m\al} - E_{n\bt} \; , \qquad  \om_{nn}^{\al\al} = 0 \; .
\ee
Therefore
\be
\label{6.17}
\rho_{mn}(t) = \sum_\al \rho_{mn}^{\al\al}(0) \exp(-i \om_{mn}^{\al\al} t )\; , \qquad
 \rho^{\al\bt}(t) = \sum_n \rho_{nn}^{\al\bt}(0) \exp(-i \om_{nn}^{\al\bt} t )\; .
\ee
Notice that the diagonal elements do not depend on time, 
$$
\rho_{nn}(t) =  \rho_{nn}(0) \; , \qquad \rho^{\al\al}(t) =  \rho^{\al\al}(0) \; .
$$

Let us introduce the distributions of states
\be
\label{6.18}
 g_{mn}(\om) = 
\sum_\al \; \frac{\rho_{mn}^{\al\al}(0)}{\rho_{mn}(0)} \; \dlt(\om-\om_{mn}^{\al\al} )\; , 
\qquad
 g^{\al\bt}(\om) = 
\sum_n \; \frac{\rho_{nn}^{\al\bt}(0)}{\rho^{\al\bt}(0)} \; \dlt(\om-\om_{nn}^{\al\bt} ) \;  ,
\ee
whose diagonal parts are
$$
 g_{nn}(\om) = g^{\al\al}(\om) = \dlt(\om) \;  .
$$
These distributions are the densities of states normalized so that 
\be
\label{6.19}
 \int_{-\infty}^\infty g_{mn}(\om) \; d\om = 1 \; , \qquad
 \int_{-\infty}^\infty g^{\al\bt}(\om) \; d\om = 1 \; .
\ee
Then the matrix elements (\ref{6.17}) can be written as
\be
\label{6.20}
\rho_{mn}(t) = \rho_{mn}(0) D_{mn}(t) \; , \qquad
 \rho^{\al\bt}(t) = \rho^{\al\bt}(0) D^{\al\bt}(t) \;  ,
\ee
with the notation
\be
\label{6.21}
 D_{mn}(t) = \int_{-\infty}^\infty g_{mn}(\om) e^{-i\om t} \; d\om \; , \qquad
 D^{\al\bt}(t) = \int_{-\infty}^\infty g^{\al\bt}(\om) e^{-i\om t} \; d\om .
\ee
Factors (\ref{6.21}) enjoy the properties
$$
D_{mn}(0) = D^{\al\bt}(0) = 1 \; , \qquad  D_{nn}(t) = D^{\al\al}(t) = 1 \; .
$$

To proceed further, let us assume that the system is sufficiently large, so that 
the state distributions $g_{mn}$ and $g^{\alpha \beta}$ are measurable, similarly
to the density of states of macroscopic systems \cite{Kittel_98}. And by definition 
(\ref{6.19}) these functions are integrable. Then by Riemann-Lebesgue lemma 
\cite{Bochner_99}, one has
$$
\lim_{t\ra\infty} D_{mn}(t) = 0 \qquad ( m\neq n) \; , 
$$
\be
\label{6.22}
 \lim_{t\ra\infty} D^{\al\bt}(t) = 0 \qquad ( \al \neq \bt) \;  .
\ee
Therefore
\be
\label{6.23}
\lim_{t\ra\infty} \rho_{mn}(t) = \dlt_{mn} \rho_{mn}(0) \; ,
\qquad
 \lim_{t\ra\infty} \rho^{\al\bt}(t) = \dlt_{\al\bt} \rho^{\al\bt}(0) \;  .
\ee
Hence in the expressions
$$
|| \hat\rho_A(t) ||^2 = 
\sum_n | \rho_{nn}(0) |^2 + \sum_{m\neq n} | \rho_{mn}(0) D_{mn}(t) |^2 \; ,
$$
\be
\label{6.24}
|| \hat\rho_B(t) ||^2 = 
\sum_\al | \rho^{\al\al}(0) |^2 + \sum_{\al\neq \bt} | \rho^{\al\bt}(0) D^{\al\bt}(t) |^2 \; ,
\ee
the nondiadonal parts tend to zero with increasing time.

In that way, measure (\ref{6.7}) varies from the initial value
\be
\label{6.25}
\ep(\hat\rho(0) ) = \frac{1}{2} \; \log \; 
\frac{||\hat\rho(0)||^2}{\sum_{mn} |\rho_{mn}(0)|^2\; \sum_{\al\bt} |\rho^{\al\bt}(0)|^2} 
\ee
to the final value
\be
\label{6.26}
 \ep(\hat\rho(\infty) ) = \frac{1}{2} \; 
\log \; \frac{||\hat\rho(0)||^2}{\sum_n |\rho_{nn}(0)|^2\; \sum_\al |\rho^{\al\al}(0)|^2}  .
\ee
From here it follows that the effect of decoherence leads to the increase of measure (\ref{6.7}),
since
\be
\label{6.27}
 \ep(\hat\rho(\infty) ) > \ep(\hat\rho(0) ) \;  .
\ee
 
As an example, illustrating how the decoherence factor $D_{mn}$ tends to zero, we may take 
the typical Lorentz form of the distribution 
$$
g_{mn}(\om) = \frac{\Gm_{mn}}{\pi(\om^2 +\Gm_{mn}^2)} \;    .
$$
Then
$$
 D_{mn}(t) = \exp(-\Gm_{mn} t ) \;  .
$$

The increase of the entanglement production measure, as is explained in Sec. 5.4, means 
that the difference, measured by the R\'{e}nyi entropy, of the system state $\hat{\rho}(t)$ 
from the nonentangling product state $\hat{\rho}_\otimes(t)$ increases under decoherence. 
In other words, the growing entanglement production measure implies that the system state
becomes more entangling as a result of decoherence.

\section{Two-qubit register in thermal bath}

As an example of an equilibrium Gibbs state, let us consider the Gibbs state of a 
two-qubit register in thermal bath. Such states are often met in quantum information 
theory. The Gibbs state is defined in the usual way, as in Eq. (\ref{26}), where the 
influence of the thermal bath is characterized by the bath temperature. Note that
this description is equivalent to the method, when one models a system-bath interaction, 
after which one averages out the bath degrees of freedom, under the assumption of thermal 
contact between the system and the bath \cite{Kubo_81,Klimontovich_82,Bogolubov_83}, 
which is effectively represented by the statistical operator of the Gibbs state defined 
in Eq. (\ref{26}), depending on the bath inverse temperature $\bt$.

\subsection{Calculating entanglement-production measure}

The system Hamiltonian is a sum
\be
\label{41}
  H = H_0 + H_{int} 
\ee
of a noninteracting part $H_0$ and an interaction term $H_{int}$. The noninteracting 
part has the Zeeman form
\be
\label{42}
 H_0 = - B \left ( S_1^z \bigotimes \hat 1_2 + \hat 1_1 \bigotimes S_2^z \right ) \;  ,
\ee
where $S_i^z$ are spin $1/2$ operators and $B$ plays the role of an external field.  
The interaction term
\be
\label{43}
 H_{int}  = - 2J S_1^z \bigotimes S_2^z
\ee
describes the qubit coupling. When $J > 0$, the coupling is called ferromagnetic,
while if $J < 0$, it is named antiferromagnetic. The Hamiltonian acts on the Hilbert space
$\cH$ being the closed linear envelope over the basis formed by the Hamiltonian 
eigenfunctions. 

Since $H_0$ and $H_{int}$ commute, one has
\be
\label{A35}
 e^{-\bt H} = e^{-\bt H_0} e^{-\bt H_{int} } \;  .
\ee
The exponential operators can be reduced to non-exponential forms \cite{Bernstein_64}.
Noticing that
$$
 H_0^{2n} = \frac{H_0^2}{B^2} \; B^{2n} \qquad ( n = 1,2,\ldots ) \;  ,
$$
\be
\label{A36}
 H_0^{2n+1} = \frac{H_0}{B} \; B^{2n+1} \qquad ( n = 0,1,2,\ldots ) \;  ,
\ee
where
\be
\label{44}
H_0^2 = \frac{B^2}{2} \; \left ( \hat 1_\cH + 4 S_1^z \bigotimes S_2^z \right ) \;   ,
\ee
we find
\be
\label{45}
 e^{-\bt H_0} = 1 + \frac{H_0^2}{B^2} \; [ \cosh(\bt B) - 1 ] - 
\frac{H_0}{B}\;\sinh (\bt B) \; .
\ee

Respectively, taking into account the relations, 
$$
H_{int}^{2n} = \left ( \frac{J}{2} \right )^{2n} \qquad ( n = 1,2,\ldots ) \;   ,
$$
\be
\label{A37}
H_{int}^{2n+1} = \left ( \frac{J}{2} \right )^{2n} H_{int} \qquad ( n = 0,1,2,\ldots ) \;   ,
\ee
we get
\be
\label{46}
 e^{-\bt H_{int} } = \cosh \left ( \frac{\bt J}{2} \right ) - 
2\; \frac{H_{int}}{J} \sinh \left ( \frac{\bt J}{2} \right ) \; .
\ee  
Combining Eqs. (\ref{45}) and  (\ref{46}) yields
$$
e^{-\bt H } =  \left \{ \frac{H_0^2}{B^2}\; [ \cosh(\bt B) - 1 ] + 
\hat 1_\cH \right \}
\cosh \left ( \frac{\bt J}{2} \right ) +
$$
\be
\label{47}
 + \left \{ \frac{H_0^2}{B^2} \; [ \cosh(\bt B) - 1 ]
- 2 \; \frac{H_{int}}{J} \right \} \sinh \left ( \frac{\bt J}{2} \right ) -
  \frac{H_0}{B} \; \sinh(\bt B) \left [ \cosh \left ( \frac{\bt J}{2} \right ) + 
\sinh \left ( \frac{\bt J}{2} \right )   \right ]  \; .
\ee
Then the partition function becomes
\be
\label{48}
Z \equiv {\rm Tr}_\cH e^{-\bt H} = 
2 [ \cosh(\bt B) + 1 ] \cosh \left ( \frac{\bt J}{2} \right )  + 
2 [ \cosh(\bt B) - 1 ] \sinh \left ( \frac{\bt J}{2} \right ) \; .
\ee
And for expression (\ref{28}) we obtain
\be
\label{49}
 f_1 \equiv || e^{-\bt H} ||^2 =  2 [ \cosh(2\bt B) + 1 ] \cosh(\bt J) +
2 [ \cosh(2\bt B) - 1 ] \sinh(\bt J) \; .
\ee

Taking the trace over $\mathcal{H}$, except $\mathcal{H}_i$, gives
\be
\label{A38}
{\rm Tr}_{\cH/\cH_i} e^{-\bt H} = \frac{1}{2}\; Z \hat 1_i -
2S_i^z \sinh(\bt B) \left [ \cosh \left ( \frac{\bt J}{2} \right ) +
\sinh \left ( \frac{\bt J}{2} \right ) \right ] \;  ,
\ee
from where
\be
\label{A39}
 {\rm Tr}_{\cH_i} \left ( {\rm Tr}_{\cH/\cH_i} e^{-\bt H} \right )^2 =
4\cosh(\bt B) + 2 [ \cosh(2\bt B) + 1 ] \cosh(\bt J) +
 2 [ \cosh(2\bt B) - 1 ] \sinh(\bt J) \; .
\ee
Thus we come to function (\ref{29}) in the form
\be
\label{50}
 f_2 = [f_1 + 4\cosh(\bt B) ]^2 \;  .
\ee

The entanglement production measure (\ref{30}) takes the form
\be
\label{51}
  \ep(\hat\rho) = \frac{1}{2} \; \log\left ( \frac{f_1}{f_2} \; Z^2 \right ) \; ,
\ee
in which
\be
\label{52}
 Z^2 = 8 \cosh(\bt B) +  4[ \cosh^2(\bt B) + 1 ] \cosh(\bt J) +
 4 [ \cosh^2(\bt B) - 1 ] \sinh(\bt J) \;   .
\ee

When the qubits are not coupled, so that $J \ra 0$, but $B \neq 0$, then
\be
\label{A40}
 f_1 \simeq 4\cosh^2(\bt B) \; , \qquad 
f_2 \simeq 16\cosh^2(\bt B) [ 1 + \cosh(\bt B)]^2 \; , \qquad 
Z^2 \simeq 4 [ 1 + \cosh(\bt B) ]^2 \;  .
\ee
And there is no entanglement production:
\be
\label{53}
\ep(\hat\rho) = 0 \qquad ( J = 0 \; , \;\; B\neq 0 ) \;   .
\ee
In the opposite case, when $B \ra 0$, but $J \neq 0$, we have
\be
\label{A41}
 f_1 \simeq 4\cosh(\bt J) \; , \qquad 
f_2 \simeq 16 [ 1 + \cosh(\bt J)]^2 \; , \qquad 
Z^2 \simeq 48[ 1 + \cosh(\bt J) ] \;   .
\ee
Then the measure is finite,
\be
\label{54}
 \ep(\hat\rho) = \frac{1}{2} \;\log \; \frac{2\cosh(\bt J)}{1+\cosh(\bt J)} \qquad
(B = 0\; , \;\; J \neq 0 ) \;  .
\ee
In the limiting case of strong coupling, it tends to the limit
\be
\label{55}
 \ep(\hat\rho) = \frac{1}{2}\;\log 2 \qquad ( B = 0 \; , \;\; J \ra \pm\infty) \;  .
\ee

Functions (\ref{49}), (\ref{50}), and (\ref{52}), defining measure (\ref{51}), are
even with respect to $B$, hence it is sufficient to consider only one sign of $B$. 
In what follows, we assume that $B$ is positive, $B > 0$. A more detailed 
analysis of the entanglement production measure (\ref{51}) should 
be done separately for the ferromagnetic and antiferromagnetic coupling.

\subsection{Entanglement production under ferromagnetic coupling ($J > 0$)}

According to Eq. (\ref{53}), there is no entanglement production without qubit 
coupling. Nontrivial behavior of measure (\ref{51}) exists only for $J \neq 0$.
It is therefore convenient to introduce the dimensionless variables
\be
\label{56}
  T \equiv \frac{1}{\bt | J |} \; , \qquad h \equiv \frac{B}{|J| } \; ,
\ee
so that measure (\ref{51}) becomes a function of these variables,
\be
\label{57}
 \ep(\hat\rho) = \ep(T,h) \; .
\ee
In the definition of measure (\ref{51}), for concreteness, we take the natural 
logarithm. The asymptotic behavior of the measure is as follows. 

At low temperature, but finite $h$, we have
\be
\label{58}
 \ep(T,h) \simeq e^{-2h/T} - e^{-4h/T} + \frac{1}{3} \; e^{-6h/T} \qquad
(T \ra 0 , \;\; h > 0 ) \;  .
\ee
In the opposite regime of small $h$, but finite temperature, we get
\be
\label{59}
\ep(T,h) \simeq a_0 + a_2 h^2 \qquad (h\ra 0, \;\; T > 0 ) \; ,
\ee
where the coefficients are
\be
\label{A42}
 a_0 = \frac{1}{2}\;\ln\;\frac{2\cosh(1/T)}{1+\cosh(1/T)} \; , \qquad
a_2 = \frac{e^{1/T}(e^{1/T}-1)(1+2e^{1/T} - e^{2/T})}{2T^2(e^{1/T}+1)^2(e^{2/T}+1)} \;  .
\ee
These expansions show that the limits of $h \ra 0$ and $T \ra 0$ are not 
commutative, since
\be
\label{60}
 \lim_{h\ra 0}\;\lim_{T\ra 0} \ep(T,h) = 0 \;  ,
\ee
while
\be
\label{61} 
  \lim_{T\ra 0}\;\lim_{h\ra 0} \ep(T,h) = \frac{1}{2}\; \ln 2 \; .
\ee

At high temperature, but finite $h$, the measure is
\be
\label{62}
 \ep(T,h) \simeq \frac{1}{8} \; \left ( e^{1/T} -1 \right )^2 +
\frac{h^2-1}{8} \; \left ( e^{1/T} -1 \right )^3 \qquad 
( T \ra \infty, \;\; h > 0 )\;  ,
\ee
which shows that
\be
\label{63}
  \lim_{T\ra \infty} \ep(T,h) = 0 \qquad ( 0 < h < \infty) \;  .
\ee
And for large $h$, but finite temperature, we find
\be
\label{64}
 \ep(T,h) \simeq  b_2 e^{-2h/T} + b_3 e^{-3h/T} \qquad 
( h \ra \infty, \;\; T > 0 )\;  ,
\ee
where
\be
\label{A43}
 b_2 = 1 - e^{-2/T} \; , \qquad b_3 = - 4e^{-1/T} \left ( 1 - e^{-2/T} \right ) \;  .
\ee
Hence 
\be
\label{65}
\lim_{h\ra \infty} \ep(T,h) = 0 \qquad ( 0 < T < \infty) \;   .
\ee

The general behavior of the entanglement production measure, under 
ferromagnetic coupling, as a function of the dimensionless variables $T$ 
and $h$, is demonstrated in Fig. 1. The maximal value of the measure 
\be
\label{A44}
 \max \; \ep(T,h) = \frac{1}{2} \; \ln 2 = 0.347
\ee
is reached when, first, $h \ra 0$, under finite $T$, after which $T \ra 0$. 
 
%Figure 1
\begin{figure}[ht]
\centerline{\includegraphics[width=7.5cm]{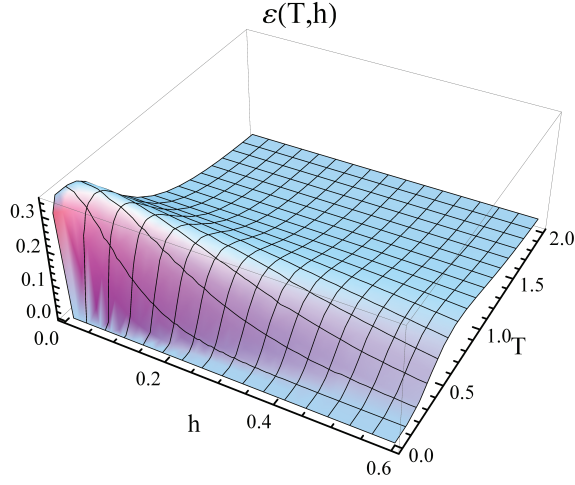}}
\caption{Measure of entanglement production, under ferromagnetic coupling
($J>0$), as a function of dimensionless variables $T$ and $h$.}
\label{fig:Fig.1}
\end{figure}

\subsection{Entanglement production under antiferromagnetic coupling ($J < 0$)}

Under antiferromagnetic coupling of qubits, the measure of entanglement production
behaves in a different way, depending on whether $h < 1$, $h = 1$, or $h > 1$. 

At low temperature and $h$ in the interval $0 \leq h < 1$ the asymptotic behavior
of the measure is
\be
\label{66}
 \ep(T,h) \simeq \frac{1}{2}\; \ln 2 \; - \; \frac{1}{2}\; e^{-(1-h)/T} \qquad
(T \ra 0 \; , \;\; 0 \leq h < 1 ) \;  ,
\ee
so that
\be
\label{67}
 \lim_{T\ra 0} \ep(T,h) = \frac{1}{2}\; \ln 2 \qquad ( 0 \leq h < 1 ) \;  .
\ee

But, if $h = 1$ and $T \ra 0$, then
\be
\label{68}
 \ep(T,h) \simeq \frac{1}{2}\; \ln \; \frac{27}{25} \; - \; \frac{1}{15}\; e^{-2/T} \; -
\; \frac{2}{225} \; e^{-4/T} \qquad (T \ra 0 \; , \;\;  h \equiv 1 ) \;  ,
\ee
which gives
\be
\label{69}
 \lim_{T\ra 0} \ep(T,h) = \frac{1}{2}\; \ln \; \frac{27}{25} \qquad ( h \equiv 1) \;  .
\ee
And, if $h > 1$, the limit of low temperatures becomes
\be
\label{70}
 \lim_{T\ra 0} \ep(T,h) = 0 \qquad ( h > 1 ) \;  .
\ee

For small $h$, but finite temperature, we have
\be
\label{71}
 \ep(T,h) \simeq c_0 + c_2 h^2 \qquad ( h \ra 0 \; , \;\; T > 0 ) \;  ,
\ee
with the coefficients
\be
\label{A45}
c_0 = \frac{1}{2}\;\ln\; \frac{2\cosh(1/T)}{1+\cosh(1/T)} \; , \qquad
c_2 = \frac{(e^{1/T}-1)(1-2e^{1/T}-2e^{2/T})}{2T^2(e^{1/T}+1)^2(e^{2/T}+1)} \;   .
\ee

At high temperature, but finite $h$, we find
\be
\label{72}
 \ep(T,h) \simeq \frac{1}{8} \left ( e^{1/T}-1 \right )^2 \; -\; \frac{h^2+1}{8}\;
 \left ( e^{1/T}-1 \right )^3 \qquad (T \ra \infty \; , \;\; h \geq 0 ) \; ,
\ee
hence
\be
\label{73}
 \lim_{T\ra \infty} \ep(T,h) = 0 \qquad ( h \geq 0 ) \;  .
\ee

And when $h \ra \infty$, at finite temperature, we obtain
\be
\label{74}
\ep(T,h) \simeq b_2 e^{-2h/T} + b_3 e^{-3h/T} \qquad 
(h \ra \infty \; , \;\; T \geq 0 ) \;   ,
\ee
with the same coefficients $b_2$ and $b_3$ as in the high-field limit (\ref{64}). 
Therefore
\be
\label{75}
 \lim_{h\ra \infty} \ep(T,h) = 0 \qquad ( T \geq 0 ) \;  .
\ee

Figure 2 shows the general behavior of the entanglement production measure, under 
antiferromagnetic coupling, as a function of the dimensionless variables $T$ and $h$. 
The maximal value (\ref{67}) is reached at low temperature and $h < 1$.

%Figure 2
\begin{figure}[ht]
\centerline{\includegraphics[width=7.5cm]{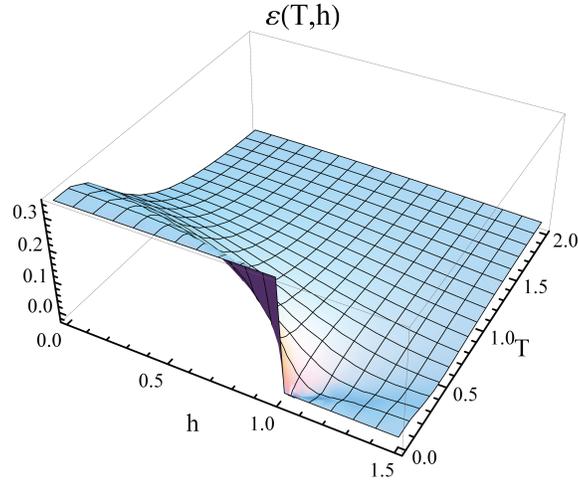}}
\caption{Measure of entanglement production, under antiferromagnetic coupling
($J<0$), as a function of dimensionless variables $T$ and $h$.}
\label{fig:Fig.2}
\end{figure}

\section{Hilbert space partitioning}

For many systems, as studied in the previous sections, the partitioning of  the Hilbert 
space has been uniquely fixed. For more complex systems, the type of partitioning 
may be not unique. Respectively, the entangling properties of the system statistical 
operator depend on which parts of the system are being entangled. To illustrate how 
different kinds of partitioning could arise, let us consider a system of $N$ particles 
with spins. For brevity, we can combine the spatial, $r_i$, and spin, $\sigma_i$,  
degrees of freedom in the notation $x_i = \{r_i, \sigma_i \}$. The system wave 
function $|\Psi_{nl}\rangle$ depends on the multi-indices $n$ and $l$ for the spatial 
and spin states, respectively. The function
\be
\label{A46}
 | \Psi_{nl} \rgl = [ \Psi_{nl}(x_1,x_2,\ldots,x_N) ]
\ee
can be treated as a column with respect to all its variables, so that its normalization
reads as
\be
\label{A47}
\lgl  \Psi_{nl} | \Psi_{nl} \rgl = 
\int | \Psi_{nl}(x_1,x_2,\ldots,x_N) |^2 \; dx_1 \ldots dx_N = 1 \;   .
\ee
As usual, summation with respect to discrete indices is assumed. The system statistical
operator
\be
\label{76}
\hat\rho_{nl} = | \Psi_{nl} \rgl \lgl  \Psi_{nl} |
\ee
acts on the Hilbert space $\mathcal{H}$.

\subsection{Particle partitioning}

The natural partitioning of the system Hilbert space is with respect to particles
composing the system. Then we can define the real-space single-particle Hilbert
space
\be
\label{77}
\cH_i^{spat} = {\rm span} \{ \vp_{n_i}(\br_i) \} 
\ee
as a closed linear envelope over a single-particle basis depending on real-space
coordinates. Similarly, a spin-dependent basis defines the Hilbert space
\be
\label{78}
 \cH_i^{spin} = {\rm span} \{ | \sigma_i \rgl  \} \;  .
\ee
Then a single-particle Hilbert space is
\be
\label{79}
 \cH_i =  \cH_i^{spat} \bigotimes \cH_i^{spin} \;  .
\ee
The total system Hilbert space can be represented as a tensor product
\be
\label{80}
\cH = \bigotimes_{i=1}^N \cH_i
\ee
of single-particle spaces.

Following the general scheme, we define the reduced statistical operators
\be
\label{81}
 \hat\rho_{n l}^{(i)} \equiv {\rm Tr}_{\cH/\cH_i} \hat\rho_{nl} \;  ,
\ee
whose tensor product induces the nonentangling operator
\be
\label{82}
 \hat\rho_{nl}^\otimes = \bigotimes_{i=1}^N \hat\rho_{n l}^{(i)} \;  .
\ee
Then the entanglement production of the statistical operator (\ref{76}), with respect
to the Hilbert space partitioning (\ref{80}), is quantified by the measure
\be
\label{83}
 \ep ( \hat\rho_{nl} ) \equiv 
\log \; \frac{||\hat\rho_{nl}||}{||\hat\rho_{nl}^\otimes||} = 
- \log || \hat\rho_{nl}^\otimes || \; .
\ee
Keeping in mind indistinguishable particles, we get
\be
\label{A48}
 ||\hat\rho_{nl}^\otimes|| = \prod_{i=1}^N ||\hat\rho_{n l}^{(i)}||   = 
||\hat\rho_{n l}^{(i)}||^N \; .
\ee
Therefore measure (\ref{83}) becomes
\be
\label{84}
 \ep(\hat\rho_{nl}) = -\; \frac{N}{2} \;\log \left ( 
{\rm Tr}_{\cH_i} (\hat\rho_{n l}^{(i)})^2\right ) \;  .
\ee

Note that we have no problems dealing with indistinguishable particles, while the
definition of state entanglement for indistinguishable particles confronts some
problems \cite{Amico_65,Benatti_66}. All we need is to correctly symmetrize the
system wave function depending on whether bosons or fermions are considered.

\subsection{Spin-spatial partitioning}

It is also interesting to study the entanglement between spin and spatial degrees
of freedom \cite{Omar_67,Karlsson_68,Lamata_69,Wang_70,Kastner_71}. To
consider such a spin-spatial entanglement production, it is necessary to partition
the system Hilbert space onto spin and spatial degrees of freedom. For this
purpose, we introduce the real-space Hilbert part
\be
\label{85}
\cH_{spat} \equiv \bigotimes_{i=1}^N \cH_i^{spat}
\ee
and the spin Hilbert space
\be
\label{86}
\cH_{spin} \equiv \bigotimes_{i=1}^N \cH_i^{spin} \;  .
\ee
Then the total Hilbert space is a tensor product of the spatial and spin parts
\be
\label{87}
\cH = \cH_{spat} \bigotimes \cH_{spin}  \; .
\ee

The related reduced operators are
\be
\label{88}
\hat \rho_{nl}^{spat} = {\rm Tr}_{\cH_{spin}} \hat\rho_{nl} \; , \qquad
\hat\rho_{nl}^{spin} = {\rm Tr}_{\cH_{spat}} \hat\rho_{nl} \;  ,
\ee
defining the non-entangling operator
\be
\label{A49}
\hat\rho_{nl}^\otimes = \hat \rho_{nl}^{spat} \bigotimes \hat\rho_{nl}^{spin} \;  .
\ee
This gives the entanglement production measure for the system statistical operator,
with respect to the entanglement of spin and spatial degrees of freedom, as
\be
\label{89}
 \overline\ep(\hat\rho_{nl}) \equiv 
\log\; \frac{||\hat\rho_{nl}||}{||\hat\rho_{nl}^\otimes||} = -
\log \left ( || \hat\rho_{nl}^{spat}|| \cdot  || \hat\rho_{nl}^{spin}|| \right )   \;  .
\ee
Employing the Hilbert-Schmidt norm yields
\be
\label{90}
   \overline\ep(\hat\rho_{nl}) =-\frac{1}{2} \left [ \log\; {\rm Tr}_{\cH_{spat}}
\left (\hat\rho_{nl}^{spat} \right )^2 + \log\; {\rm Tr}_{\cH_{spin}}
\left (\hat\rho_{nl}^{spin} \right )^2 \right ] \; .
\ee
Quantities (\ref{84}) and (\ref{90}) are different. In the following section, we
present explicit calculation of their values.

\section{Multiparticle spinor quantum system}

\subsection{Permutation-invariant wavefunctions of spinor particles}

This section formulates general properties of many-body wavefunctions of indistinguishable 
spinor particles with separable spin and spatial degrees of freedom. Such a system is 
described by the Hamiltonian $\hat{H}_{\mathrm{spat}}+\hat{H}_{\mathrm{spin}}$, where 
$\hat{H}_{\mathrm{spat}}$ is spin-independent, $\hat{H}_{\mathrm{spin}}$ is 
spatially-homogeneous, and each of $\hat{H}_{\mathrm{spat}}$ and $\hat{H}_{\mathrm{spin}}$ 
is permutation-invariant. The wavefunctions are composed from the spin $\Xi_{tl}^{[\lambda]}$ 
and spatial $\Phi_{tn}^{[\lambda]}$ functions, which form bases of irreducible representations 
of the symmetric group $\pr S_{N}$ of $N$-symbol permutations 
(see \cite{hamermesh,Kaplan,pauncz_symmetric,yurovsky14,yurovsky15}). This means 
that a permutation $\pr P$ of the particles transforms each basis function to a linear combination
of the functions in the same representation, 
\begin{align}
\pr P\Xi_{tl}^{[\lambda]}  & =  \sum_{t'}D_{t't}^{[\lambda]}(\pr P)\Xi_{t'l}^{[\lambda]} \; ,
\\
\pr P\Phi_{tn}^{[\lambda]} & = \mathrm{sgn}(\pr{P}) \sum_{t'}D_{t't}^{[\lambda]}(\pr P)\Phi_{t'n}^{[\lambda]} \; .
\end{align}
Here, the irreducible representations are associated with the Young diagram 
$\lambda=[\lambda_1,\ldots,\lambda_M]$. The number of the diagram rows $M=2s+1$ is 
the multiplicity, where $s$ is the particle's spin. The basic functions of the representation 
are labeled by the standard Young tableaux $t$ of the shape $\lambda$. The factor 
$\mathrm{sgn}(\pr P)$ is the permutation parity for fermions and $\mathrm{sgn}(\pr P)\equiv 1$ 
for bosons.  The Young orthogonal representation matrices $D_{t't}^{[\lambda]}(\pr P)$ satisfy 
the following relations,
\begin{align}
 D_{t t'}^{[\lambda]}(\pr{P})&=D_{t' t}^{[\lambda]}(\pr{P}^{-1}) \; ,
 \label{InvPerm}
 \\
 \sum_{t'} D_{r t'}^{[\lambda]}(\pr{P}) D_{t' t}^{[\lambda]}(\pr{Q})
 &= D_{r t}^{[\lambda]}(\pr{P}\pr{Q}) \; ,
 \label{ProdYoung}
 \\
 \sum_{\pr{P}} D_{t' r'}^{[\lambda']}(\pr{P})D_{t r}^{[\lambda]}(\pr{P})
&= \frac{N!}{f_{\lambda}}\delta_{\lambda \lambda'} \delta_{t t'} \delta_{r r'}  \; ,
 \\
 D_{t' t}^{[\lambda]}(\pr{E})&=\delta_{t' t} \; ,
 \label{IdentPerm}
\end{align}
where $\pr{E}$ is the identity permutation. These relations provide the proper bosonic 
or fermionic permutation symmetry of the total wavefunction
\begin{equation}
\Psi_{nl}^{[\lambda]}=f_{\lambda}^{-1/2}\sum_{t}\Phi_{tn}^{[\lambda]}\Xi_{tl}^{[\lambda]} \; ,
\label{Psilamnl}
\end{equation}
$\pr{P}\Psi^{[\lambda]}_{n l}=\mathrm{sgn}(\pr{P})\Psi^{[\lambda]}_{n l}$. The representation 
dimension is given by
\begin{equation}
f_{\lambda}=\frac{N!\prod_{m<m'}(\lambda_m-m-\lambda_{m'}+m')}
 {\prod_{m=1}^M (\lambda_m+M-m)!} \; .
\label{flambda}
\end{equation}

For spin-$1/2$ particles, the Young diagrams have two rows and are unambiguously 
related to the total spin $S$, $\lambda=[N/2+S,N/2-S]$. The representation dimension 
can be expressed as
\begin{equation}
f_{\lambda}=\frac{N!(2S+1)}{(N/2+S+1)!(N/2-S)!} \label{flambda2} \; .
\end{equation}

An explicit expression for the spin wavefunction is obtained \cite{yurovsky13} in the case 
of commutative $\hat{H}_{\mathrm{spin}}$ and the total spin projection operator $\hat{S}_{z}$, 
\begin{equation}
\Xi_{tS_{z}}^{[\lambda]}=C_{SS_{z}}\sum_{\pr P}D_{t[0]}^{[\lambda]}(\pr P)\prod_{j=1}^{N/2+S_{z}}|\uparrow(\pr Pj)\rangle\prod_{j=N/2+S_{z}+1}^{N}|\downarrow(\pr Pj)\rangle  \label{XiStSz}  \; .
\end{equation}
The wavefunction is unambiguously determined by the total spin $S$ and its projection
$S_{z}$, which is the half of the difference of the occupations of the two spin states 
$|\uparrow\rangle$ and $|\downarrow\rangle$.
The normalization factor is expressed as \cite{yurovsky13}
\begin{equation}
C_{SS_{z}}=\frac{1}{(N/2+S_{z})!(N/2-S)!}\sqrt{\frac{(2S+1)(S+S_{z})!}{(N/2+S+1)(2S)!(S-S_{z})!}} \label{CSSz}  \; .
\end{equation}

\subsection{Spin-spatial partitioning}

The spin-spatial entanglement production measure can be evaluated for a generic system 
of  indistinguishable spinor particles with separable spin and spatial degrees of freedom. 
Due to the orthogonality of the spin and spatial functions, 
$\left\langle \Xi_{t'l}^{[\lambda]}|\Xi_{tl}^{[\lambda]}\right\rangle =\delta_{tt'}$,
$\left\langle \Phi_{tn}^{[\lambda]}|\Phi_{t'n}^{[\lambda]}\right\rangle =\delta_{tt'}$,
we have for the total wavefunction (\ref{Psilamnl})
\begin{equation}
\hat{\rho}^{\mathrm{spin}}_{n l}={\rm Tr}_{\cH_{spat}}\left|\Psi_{nl}^{[\lambda]}\right\rangle \left\langle \Psi_{nl}^{[\lambda]}\right|=\frac{1}{f_{\lambda}}\sum_{t,t'}\left\langle \Phi_{tn}^{[\lambda]}|\Phi_{t'n}^{[\lambda]}\right\rangle \left|\Xi_{tl}^{[\lambda]}\right\rangle \left\langle \Xi_{t'l}^{[\lambda]}\right|=\frac{1}{f_{\lambda}}\sum_{t}\left|\Xi_{tl}^{[\lambda]}\right\rangle \left\langle \Xi_{tl}^{[\lambda]}\right| 
\end{equation}
and, similarly,
\begin{equation}
\hat{\rho}^{\mathrm{spat}}_{n l}=
\frac{1}{f_{\lambda}}\sum_{t}\left|\Phi_{tn}^{[\lambda]}\right\rangle \left\langle \Phi_{tn}^{[\lambda]}\right| \; .
\end{equation}
Then
\begin{equation}
{\rm Tr}_{\cH_{spat}}\left(\hat{\rho}^{\mathrm{spat}}_{n l}\right)^{2}=\frac{1}{f_{\lambda}^{2}}\sum_{t,t'}\left\langle \Phi_{tn}^{[\lambda]}|\Phi_{t'n}^{[\lambda]}\right\rangle \left\langle \Phi_{t'n}^{[\lambda]}|\Phi_{tn}^{[\lambda]}\right\rangle =\frac{1}{f_{\lambda}^{2}}\sum_{t,t'}\delta_{tt'}^{2}=\frac{1}{f_{\lambda}}
\end{equation}
and
\begin{equation}
{\rm Tr}_{\cH_{spin}}\left(\hat{\rho}^{\mathrm{spin}}_{n l}\right)^{2}=\frac{1}{f_{\lambda}} \; .
\end{equation} 
Therefore, according to Eq. (\ref{90}), the entanglement production measure  
\begin{equation}
\overline\ep(\hat\rho_{nl})=\ln f_{\lambda}
\end{equation}
depends only on the representation dimension (\ref{flambda}).

The leading term of the asymptotic expansion in the limit $N\rightarrow\infty$ can be 
evaluated using the Stirling formula in Eq. (\ref{flambda}) as
\begin{equation}
\overline\ep(\hat\rho_{nl})\sim N\ln N-\sum_{m=1}^N  \lambda_m \ln\lambda_m \; .
\end{equation} 
Its maximum, $\overline\ep(\hat\rho_{nl})= N\ln M$, is attained for equal lengths of the 
Young diagram rows $\lambda_m=N/M$.

\begin{figure}
\centerline{
\includegraphics[width=3.4in]{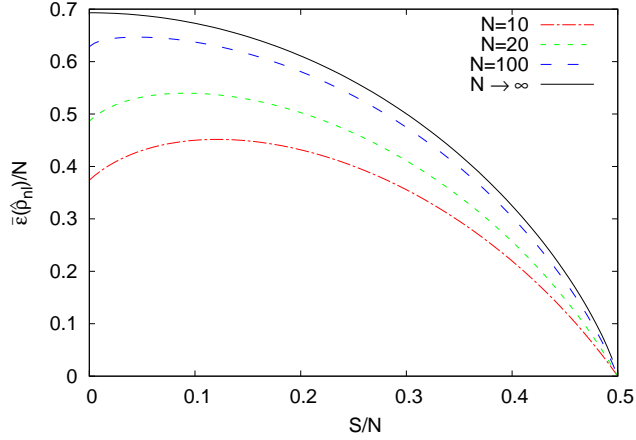}}
\caption{The spin-spatial entanglement production measure for the statistical 
operator of $N$ spin-$\frac{1}{2}$ particles in the state with the defined spin $S$. 
The red dot-dashed, green short-dashed, and blue long-dashed lines correspond to 
$N=10$, $20$, and $100$, respectively. The limiting case of $N \to \infty$ is shown by the 
black solid line.}
\label{FigSpinSpat}
\end{figure}

For spin-$1/2$ particles, the entanglement production measure decreases when $S$ 
increases (see Fig. \ref{FigSpinSpat}). The measure vanishes at $S=N/2$, when the total
wavefunction is a single product of the spin and spatial functions. The plot for $N \to \infty$ 
is obtained using the leading term in the asymptotic expansion,
\begin{equation}
\overline\ep(\hat\rho_{nl})\sim - N\left[\left( \frac{1}{2}-\frac{S}{N}\right)\ln\left( \frac{1}{2}-\frac{S}{N}\right)  
+\left(\frac{1}{2}+\frac{S}{N}\right)\ln\left( \frac{1}{2}+\frac{S}{N}\right)\right] \; .
\end{equation} 
In the asymptotic limit, the entanglement production measure attains its maximum 
of $\overline\ep(\hat\rho_{nl})= N\ln2$ at $S=0$, when the Young diagram rows have the 
equal length.

%Figure 3

\subsection{Particle partitioning}

The particle entanglement production measure can be evaluated in the particular case 
of non-interacting particles with $s=\frac{1}{2}$. If there are several spatial orbitals 
$|0\rangle,\ldots|M-1\rangle$, there are multiple spatial wavefunctions for the given $\lambda$ 
and the orbital occupations, even if $\hat{H}_{\mathrm{spat}}$ commutes
with the orbital occupations (see \cite{Kaplan,yurovsky14,yurovsky15}).
However, if there are only two spatial orbitals, $|+\rangle$ and
$|-\rangle$, and $\hat{H}_{\mathrm{spat}}$ commutes with the ``isotopic spin'' 
$\hat{I}_{z}=\sum_{j=1}^{N}(|+(j)\rangle\langle+(j)|-|-(j)\rangle\langle-(j)|)/2$,
the spatial wavefunction is unambiguously determined by the total
spin $S$ and the eigenvalue $I_{z}$ of $\hat{I}_{z}$
(it is nothing but the half of the difference of the orbital occupations) 
and can be represented for bosons like (\ref{XiStSz}),
\begin{equation}
\Phi_{t I_{z}}^{[\lambda]}=C_{S I_{z}}\sum_{\pr P}D_{t[0]}^{[\lambda]}(\pr P)\prod_{j=1}^{N/2+I_{z}}|+(\pr Pj)\rangle\prod_{j=N/2+I_{z}+1}^{N}|-(\pr Pj)\rangle  \label{PhiStSz} \; . 
\end{equation}
Here the normalization factor is defined by Eq. (\ref{CSSz}).
Given $S$, the system state is specified by two independent
spin projections, $S_{z}$, and $I_{z}$. Then the multi-indices $n$ and $l$ can be specifically
chosen as $I_{z}$ and $S_{z}$. Ground states of such systems were analyzed 
in Refs. \cite{kuklov2002,ashhab2003} using $SU(2)$ symmetry ($SU(2)$ and symmetric 
groups are closely related, having a common set of basic functions of irreducible 
representations, see \cite{Kaplan}). 

In the particular basic, the reduced statistical operators (\ref{81}) have the following 
explicit form,
\begin{equation}
\hat{\rho}^{(i)}_{I_{z} S_{z}}=
\sum_{\{n\},\{\sigma\}}\prod_{i'\neq i}\langle n(i')|\langle \sigma(i') |\Psi_{I_{z} S_{z}}^{[\lambda]} \rangle\langle\Psi_{I_{z} S_{z}}^{[\lambda]}|\prod_{i''\neq i}|n(i'')\rangle |\sigma(i'')\rangle \; , 
\end{equation}
where $n$ can be $+$ or $-$, $\sigma$ can be $\uparrow$ or $\downarrow$, 
and the summation is performed over all $n_{i'}$ and $\sigma_{i'}$ with $i'\neq i$. 
Their matrix elements can be expressed as matrix elements of the projection operator 
\begin{equation}
\langle n',\sigma'|\hat{\rho}^{(i)}_{I_{z} S_{z}}|n,\sigma\rangle
=\langle \Psi_{I_{z} S_{z}}^{[\lambda]}|n(i)\rangle |\sigma(i)\rangle\langle n(i)|\langle \sigma(i) |\Psi_{I_{z} S_{z}}^{[\lambda]}\rangle \delta_{nn'}\delta_{\sigma\sigma'}  \:  .
\label{tilrhoiSz}
\end{equation}
Due to permutation symmetry of the total wavefunction, the matrix element
\begin{equation}
\langle n,\sigma|\hat{\rho}^{(i)}_{I_{z} S_{z}}|n,\sigma\rangle=
\langle \Psi_{I_{z} S_{z}}^{[\lambda]}|\hat{P}_{n\sigma}|\Psi_{I_{z} S_{z}}^{[\lambda]}\rangle                                        
\end{equation}  
is independent of $i$. Here  
$$
\hat{P}_{n\sigma}\equiv \frac{1}{N}\sum_{i=1}^N|n(i)\rangle |\sigma(i)\rangle\langle n(i)|\langle \sigma(i) |
=
$$
\begin{equation}
= \frac{1}{4}+\frac{\delta_{\sigma \uparrow}-\delta_{\sigma \downarrow}}{2 N} \hat{S}_z
+\frac{\delta_{n +}-\delta_{n -}}{2 N}\hat{I}_{z}
+\frac{(\delta_{\sigma \uparrow}-\delta_{\sigma \downarrow})(\delta_{n +}-\delta_{n -})}{N}\sum_{i=1}^N\hat{s}_z(i)\hat{i}_{z}(i) 
\end{equation} 
is  represented in terms of the spin
$\hat{s}_{z}(i)=\frac{1}{2}(|\uparrow(i)\rangle\langle\uparrow(i)|-|\downarrow(i)\rangle\langle\downarrow(i)|)$
and isotopic spin $\hat{i}_{z}(i)=\frac{1}{2}(|+(i)\rangle\langle +(i)|-|-(i)\rangle\langle -(i)|)$ of the $i$th particle. 
The operator $\sum_{i=1}^N\hat{s}_z(i)\hat{i}_{z}(i)$, as an operator in the spin space, is a 
component of an irreducible spherical vector (see \cite{yurovsky15}). Then its matrix elements 
between states with arbitrary $S_z$ can be related to ones for $S_z=S$ using the 
Wigner-Eckart theorem (see \cite{yurovsky15}). In the case of two spacial orbitals, the same 
can be done for $I_{z}$ too, providing
\begin{equation}
 \langle \Psi_{I_{z} S_{z}}^{[\lambda]}|\sum_{i=1}^N\hat{s}_z(i)\hat{i}_{z}(i)|\Psi_{I_{z} S_{z}}^{[\lambda]}\rangle=
 \frac{I_{z} S_{z}}{S^2}\langle \Psi_{S S}^{[\lambda]}|\sum_{i=1}^N\hat{s}_z(i)\hat{i}_{z}(i)|\Psi_{S S}^{[\lambda]}\rangle \; .
\end{equation} 
Since $\Psi_{I_{z} S_{z}}^{[\lambda]}$ is an eigenfunction of $\hat{S}_z$ and $\hat{I}_{z}$, 
the matrix element of the reduced statistical operators can be related to the one for $S_z=I_{z}=S$,
$$
\langle n,\sigma|\hat{\rho}^{(i)}_{I_{z} S_{z}}|n,\sigma\rangle=
\frac{1}{4}\left(1-\frac{I_{z} S_{z}}{S^2}\right)
+\frac{\delta_{n +}-\delta_{n -}}{2 N}I_{z}\left(1-\frac{S_{z}}{S}\right) +
$$
\begin{equation}
+\frac{\delta_{\sigma \uparrow}-\delta_{\sigma \downarrow}}{2 N}S_{z}\left(1-\frac{I_{z}}{S}\right)
+\frac{I_{z} S_{z}}{S^2}\langle n,\sigma|\hat{\rho}^{(i)}_{S S}|n,\sigma\rangle \; .
\label{hatrhoSz}
\end{equation} 
The latter matrix element can be transformed, using Eqs. (\ref{Psilamnl}) and (\ref{tilrhoiSz}), 
to the sum of the products
\begin{equation}
\langle n,\sigma|\hat{\rho}^{(i)}_{S S}|n,\sigma\rangle=\frac{1}{f_{\lambda}} \sum_{t,t'}
\langle \Xi_{tS}^{[\lambda]}|\sigma(i)\rangle\langle \sigma(i) |\Xi_{t'S}^{[\lambda]}\rangle
\langle \Phi_{t S}^{[\lambda]}|n(i)\rangle\langle n(i)|\Phi_{t' S}^{[\lambda]}\rangle
\end{equation} 
of the spin and spatial matrix elements. The spin matrix elements can be represented as 
\begin{equation}
\langle \Xi_{tS}^{[\lambda]}|\sigma(i)\rangle\langle \sigma(i) |\Xi_{t'S}^{[\lambda]}\rangle
=[\delta_{\sigma \downarrow}\delta_{tt'}+(\delta_{\sigma \uparrow}-\delta_{\sigma \downarrow})
\langle\Xi_{t'S}^{[\lambda]}|\uparrow(i)\rangle\langle\uparrow(i)|\Xi_{tS}^{[\lambda]}\rangle] \; ,
\end{equation} 
where 
\begin{equation}
 \langle\Xi_{t'S}^{[\lambda]}|\uparrow(i)\rangle\langle\uparrow(i)|\Xi_{tS}^{[\lambda]}\rangle=
 (\lambda_{1}-1)!\lambda_{2}!\lambda_{1}^{2}C_{SS}^{2}
\sum_{\pr Q}D_{t[0]}^{[\lambda]}(\pr Q)D_{t'[0]}^{[\lambda]}(\pr Q)\delta_{i\pr Q\lambda_{1}}
\end{equation} 
was calculated in Ref. \cite{yurovsky15}.

Using similar expressions for the spatial matrix elements, Eqs. (\ref{InvPerm}), (\ref{ProdYoung}), 
and (\ref{IdentPerm}), one gets
$$
\langle n,\sigma|\hat{\rho}^{(i)}_{S S}|n,\sigma\rangle=
 \delta_{\sigma \downarrow}\delta_{n -} +
$$
\begin{equation}
 + (\delta_{\sigma \uparrow}\delta_{n -}+\delta_{\sigma \downarrow}\delta_{n +}
 -2\delta_{\sigma \downarrow}\delta_{n -})\frac{\lambda_1}{N}
 +(\delta_{\sigma \uparrow}-\delta_{\sigma \downarrow})(\delta_{n +}-\delta_{n -})\left( \frac{\lambda_1}{N!}\right)^2 f_{\lambda}\varSigma_{jj}^{(S,S)} \; ,
\end{equation} 
where
\begin{equation}
 \varSigma_{jj}^{(S,S)} =\frac{N!(N-1)!}{f_{S}\lambda_{1}^{2}}\left[\lambda_{1}-\frac{\lambda_{2}}{\lambda_{1}-\lambda_{2}+2}\right] 
\end{equation} 
was calculated in \cite{yurovsky15}.

\begin{figure}
\centerline{
\includegraphics[width=3.4in]{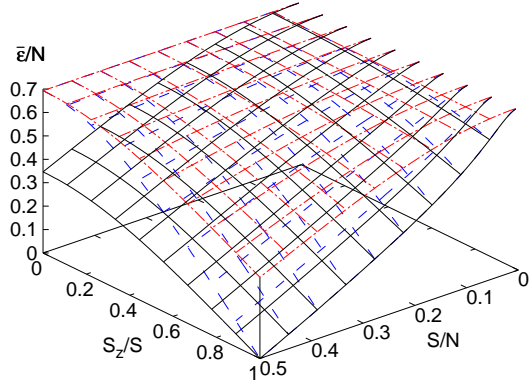} }
\caption{The particle entanglement production measure for the statistical operator 
of $N=10$ spin-$\frac{1}{2}$ particles in the state with the defined spin $S$, calculated 
with Eq. (\ref{partent}). The red dot-dashed, blue dashed, and black solid  lines correspond 
to $I_{z}=0$, $I_{z}=S_z$, and $I_{z}=S$, respectively. }
\label{FigSurfPEnt}
\end{figure}

\begin{figure}
\centerline{
\includegraphics[width=3.4in]{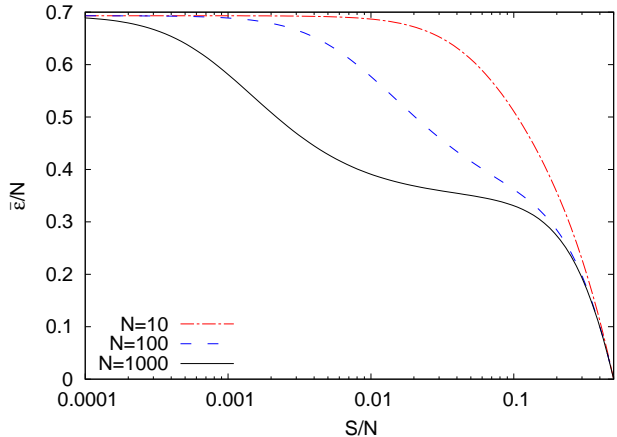} }
\caption{The particle entanglement production measure for the statistical operator of $N$ 
spin-$\frac{1}{2}$ particles in the state with $I_{z}=S_z=S$. The red dot-dashed, blue dashed, 
and black solid lines are calculated 
with Eq. (\ref{partent}) for $N=10$, $100$, and $1000$, respectively.}
\label{FigPartEnt}
\end{figure}

Then Eqs. (\ref{84}) and (\ref{hatrhoSz}) provide the particle entanglement production measure
\begin{equation}
\ep(\hat\rho_{I_{z} S_{z}})=-\frac{N}{2}\ln\left(\frac{1}{4}+\frac{1}{N^{2}}\left[S_{z}^{2}+I_{z}^{2}+\frac{(N+2)^{2}}{4S^{2}(S+1)^{2}}S_{z}^{2}I_{z}^{2}\right]\right) \; .
\label{partent} 
\end{equation}
Its maximal value $N\ln2$ is attained at $S_{z}=I_{z}=0$ for any $S$ and $N$ 
(see Figs. \ref{FigSurfPEnt} and \ref{FigPartEnt}). In the case of the spin-spatial partition, 
this value can be reached only in the limit $N \to \infty$. The particle and spin-spatial 
entanglement production measures both vanish at $S_{z}=I_{z}=S=N/2$.
However, given $0<S_{z}<N/2$ or $0<I_{z}<N/2$, the particle entanglement
increases with $S$, being maximal at $S=N/2$, when the total wavefunction
is a single product of the spin and spatial functions and the spin-spatial entanglement vanishes.

%Figure 4

%Figure 5

\section{Conclusion}

Dealing with statistical operators, one can consider two different notions. One is the 
{\it state entanglement} characterizing the structure of the given statistical operator. 
The other notion is the {\it entanglement production} by the statistical operator, 
describing the action of the statistical operator on the given Hilbert space and showing
how this action creates entangled functions from disentangled ones. These two notions
are principally different and should not be confused.   

The operational meaning of the entangling power of statistical operators is the same as 
for any other operator defined on a Hilbert space: it shows the ability of an operator
to produce entangled wave functions of the given Hilbert space. Throughout the paper,
the notion of entanglement production has been used in line the commonly accepted
in mathematical literature \cite{Marcus_13,Westwick_14,Johnston_15,Beasley_16,Alfsen_17,
Friedland_18,Gohberg_19,Crouzeux_20,Chen_23}.  

Entangling properties of statistical operators play an important role in several branches 
of quantum theory, such as quantum measurements, quantum information processing, 
quantum computing, and quantum decision theory, where one deals with composite 
measurements and composite events, related to composite probability measures. 
Entangling properties of statistical operators influence the structure of probability 
measures they generate. Depending on whether the statistical operator is entangling 
or not, the resulting probability measure can be either not factorizable or factorizable,
as is discussed in Sec. 2.  

We have defined the measure of entanglement production by statistical operators 
and illustrated it by several examples of entangled pure states, equilibrium Gibbs 
states, and by the case of a multiparticle spinor system. The relation of the introduced 
measure to other known concepts, such as quantum state purity, linear entropy or impurity, 
inverse participation ratio, quadratic R\'{e}nyi entropy, and correlators in composite
measurements, is thoroughly discussed. The measure can be defined for a collection of
quantum systems or for a set of operators characterizing a quantum system after 
measurements. The phenomenon of decoherence is also shown to be intimately related
to entanglement production.     

For complex spinor systems, the measure of entanglement production depends on the type 
of partitioning of the total Hilbert space. Thus, it is possible to realize particle partitioning 
or spin-spatial partitioning. Both these cases are analyzed. The analysis demonstrates when 
the entanglement production is maximal and when it tends to zero, which can be used 
in the applications of quantum theory mentioned above.

\section*{Acknowledgments}

This research was supported in part by a grant No. 2015616 from the United States-Israel 
Binational Science Foundation (BSF) and the United States National Science Foundation (NSF).

\end{document}